# 13

# The Great Saturn Storm of 2010-2011

**Agustín Sánchez-Lavega**[1]

**Georg Fischer**[2]

**Leigh N. Fletcher**[3]

**Enrique García-Melendo**[1, 4]

**Brigette Hesman**[5]

**Santiago Pérez-Hoyos**[1]

**Kunio M. Sayanagi**[6]

**Lawrence A. Sromovsky**[7]

1 – *Departamento Física Aplicada I, Escuela Ingeniería Bilbao, Universidad del País Vasco UPV-EHU, Bilbao (Spain).*
2 - *Space Research Institute. Austrian Academy of Sciences, Graz (Austria)*
3 – *Department of Physics and Astronomy. University of Leicester, Leicester (U. K.)*
4 - *Fundació Observatori Esteve Duran, Barcelona (Spain)*
5 – *Departament of Astronomy, University of Maryland, Maryland (U.S.A.)*
6 - *Department of Atmospheric and Planetary Sciences. Hampton University, Hampton (U.S.A.)*
7 - *Space Science and Engineering Center. University of Wisconsin, Madison (U.S.A.)*

**Abstract**

In December 2010, a major storm erupted in Saturn's northern hemisphere near 37° planetographic latitude. This rather surprising event, occurring at an unexpected latitude and time, is the sixth 'Great White Spot' (GWS) storm observed over the last century and a half. Such GWS events are extraordinary, planetary-scale atmospheric phenomena that dramatically change the typically bland appearance of the planet. Occurring while the Cassini mission was on-orbit at Saturn, the Great Storm of 2010-2011 was well-suited for intense scrutiny by the suite of sophisticated instruments onboard the Cassini spacecraft as well by modern instrumentation on ground-based telescopes and onboard the Hubble Space Telescope. This GWS erupted on December 5th close to the peak of a westward jet and generated a major dynamical disturbance that affected the whole latitude band from 25° to 48° N. At the upper cloud level, following the rapid growth of the bright outbreak spot, a blunt aerodynamic shape head formed due to interaction of the spot with the westward zonal jet, with the winds reaching velocities of 160 ms⁻¹ along the periphery of the arc. Eastward of the head, the disturbance progressed in the following months forming a turbulent wake or tail with growing vortices, one of them a major enduring anticyclone (called AV) with a size of ~ 11,000 km. Lightning events were prominent and detected as outbursts and flashes at the head and along the disturbance at both optical and radio wavelengths. The activity of the head ceased after about seven months when AV reached it, leaving the cloud structure and ambient winds perturbed. The tops of the optically dense clouds of the head reached the 300 mbar altitude level (~ 50 km below tropopause) where a mixture of ices was detected, including (1) a component of water ice lofted over 200 km altitude from its 10-bar condensation level, (2) ammonia ice as the predominant component, and (3) a component that might be ammonium hydrogen sulfide ice. The energetics of the frequency and power of lightning, as well as the estimated power generated by the latent heat released in the water-based convection to create the observed dynamical 3-dimensional flows, both indicate that the power released for much of the 7-month lifetime of the storm (~ $10^{17}$ W) was a significant fraction of Saturn's total radiated power (~ $2.2 \ 10^{17}$ W). A post storm depletion of ammonia vapor was also measured in the upper troposphere. The effects of the storm propagated into the stratosphere forming two warm airmasses at the ~ 0.5-5 mbar pressure level altitude, that later merged into a so called 'beacon' because of its 80 K temperature excess relative to its surroundings. Related to the stratospheric disturbance, hydrocarbon composition excesses were found, in particular for ethylene ($C_2H_4$), in the high stratosphere at the ~ 0.1-0.5 mbar altitude level. Numerical models of the storm dynamics explain the major observed features that essentially result from two processes: (1) A huge and sustained moist convective storm at the water clouds (altitude level 10-12 bar or ~ 250-275 km below the tropopause) and (2) the interaction of the updraft columns with the ambient winds that generates the turbulent wake consisting of vortices and waves. Model simulations of the GWS require a low vertical shear of the zonal winds and low static stability across the weather layer where the disturbance develops. Its upward propagation into the stratosphere involves Rossby waves and their breaking and energy deposition to form the beacon and induce chemical changes.
The decades-long interval between storms is probably related to the insolation cycle and the long radiative time constant of Saturn's atmosphere, and several theories for temporarily storing energy have been proposed.

## 13.1. Introduction

The Great Saturn Storm phenomenon that took place in 2010-11 is the last example of a type of huge and rare planet-encircling disturbance that occurs on this planet, a phenomenon nicknamed as a Great White Spots (GWS) event (Sanchez-Lavega, 1982). The 2010 GWS was the 6th recorded event in the last 140 years (Sanchez-Lavega, 1994; Sanchez-Lavega et al., 2011). The GWS phenomenon is easily recognizable because of the prominent brightness of the outbreak spot at visible wavelengths in a usually bland disk, and because of its size and rapid growth and expansion in longitudinal direction (Sanchez-Lavega, 1994). These disturbances have appeared in intervals of approximately 30 years at different latitudes and always in the northern hemisphere: at the equator (years 1876, 1933 and 1990), at mid-latitudes (1903, 2010) and at a subpolar latitude (1960) (Figure 13.1). Because of the 26.7°, tilt of the rotation axis of Saturn relative to its orbital plane, the North – South hemispheric visibility of the planet from the Earth changes along Saturn's 29.457-year orbital cycle. This visibility is further reduced due to the shadow of the rings on the globe and thus full hemispheric visibility occurs during short periods of about 1-2 years coincident with near-simultaneous ring-plane crossings by the Sun and Earth-based observer. This means that the GWS statistics and perceived periodicity can be biased by such sampling effects; nevertheless, it is well-established that all the recorded events have occurred in the Northern hemisphere. Thus, any conclusion of their possible seasonal origin must be regarded with caution.

**[FIGURE 13.1]**

Saturn's GWS represent a singular phenomenon in planetary atmospheres, similar in some aspects to the other planetary scale disturbances that occur in Jupiter's South Equatorial Belt (SEB Disturbances, Sanchez-Lavega and Gomez, 1996a) and temperate latitudes (NTB Disturbance Sanchez-Lavega et al, 2008). The GWS represent a class of meteorological phenomena that can potentially reveal new insights into the vigorous processes of moist convection in a hydrogen dominated atmosphere, very different from that of the Earth. As well, they allow the study of dynamic phenomena associated with the interaction of storms with a surrounding atmosphere dominated by zonal winds and its propagation along a whole latitude band, eventually encircling the planet. Finally they give clues on the poorly known structure of Saturn's atmosphere below the upper hazes and clouds, in the weather layer where thermochemical models predict active cloud formation.

In this chapter we review the observations and theoretical studies so far performed on these giant storms. We review the GWS phenomena, including the historical cases, but we mainly focus on the well-studied event of 2010-11 on which was trained the full suite of Cassini instruments as well as numerous advanced ground-based telescopic and Hubble Space Telescope instruments.

## 13.2. Review of the historical "Great White Spot" phenomenon

Until the adoption of CCD technology into ground-based telescopic observations during the mid-1980s, the detection of discrete atmospheric features in Saturn's atmosphere was a very difficult task due to their meagre contrast in the hazy atmosphere. Sanchez-Lavega (1982) summarized the small number of "spots" tracked on the planet from the time of W. Herschel (1793-94) until the Voyager 1 and 2 encounters in 1980-81. Of all these features, the most conspicuous ones were those characterized by their rapid appearance, high brightness, growth and expansion to a planetary-scale disturbance, and by their several- month lifetimes, becoming known as GWS (Great White Spot) events. To date there have been six GWS observed as summarized in Table 13.1. The data forthcoming from observations of these global-scale phenomena were scarce until the 1990 GWS. For the four previous cases (1876, 1903, 1933, 1960) details were captured mainly in drawings showing their temporal evolution except for the best case studied, i.e. that of 1933, for which there is a photographic series obtained at various times and wavelengths (Sanchez-Lavega, 1982; Sanchez-Lavega and Battaner, 1987). For the great equatorial event of 1990, CCD images were obtained from ground-based telescopes, and the Hubble Space Telescope acquired several good imaging series, making it the best-studied case until the 2010-2011 event. (Sanchez-Lavega et al., 1991 ; Barnet et al., 1992 ; Beebe et al., 1992 ; Westphal et al., 1992). Figure 13.2 shows a mosaic of the appearance of the GWS for these five historical cases.

**[TABLE 13.1]**

**[FIGURE 13.2]**

In essence, the GWS phenomenon involves three stages of development (Sánchez-Lavega, 1994): (1) The outbreak of a small but very bright "spot" that represents the onset phase; (2) the rapid growth and horizontal expansion of this "spot" into the full GWS feature, with divergence in the forming ammonia cloudtops near the ~ 1-bar level reaching ~ 10,000 – 20,000 km in size in about ten days; and (3) the planetary-scale mature disturbance phase, when the convective source, remaining active for months, interacts with the wind-sheared atmosphere. The dynamics of the mature disturbance strongly depends on the latitude of the outbreak and the meridional wind profile in its surroundings.

The source spot is brighter than its surrounding clouds over the entire visible spectrum from ~ 380 nm – 1 μm, as observed in spectral observations of the events of 1933, 1990, and 2010 (Sánchez-Lavega and Battaner, 1987; Sánchez-Lavega et al., 1991; Westphal et al., 1992; Acarreta and Sánchez-Lavega, 1999; Sánchez-Lavega et al., 2011; Sanz-Requena et al., 2012; section 13.4). The high brightness and spectral behaviour is consistent with high cloud tops formed by fresh mildly polluted ices (section 13.4). For these three cases (1933, 1990, 2010) it has been possible to measure the expanding velocity of the initial spot $V \sim$ 30-50 ms$^{-1}$ and its area growing rate with maximum values $dA/dt(max) \sim$ 200 km$^2$s$^{-1}$. Both facts suggest that the source is driven by vigorous moist convection starting at the deep water clouds at level ~ 10 bars or ~ 250 km below the visible cloud tops (Sanchez-Lavega and Battaner, 1987; Hueso and Sanchez-Lavega, 2004) (see section 13.7). This hypothesis has been recently supported by the observation of lightning events in the optical and radio wavelengths in the 2010-11 GWS (Fischer et al., 2011a; section 13.6) and the detection of water-ice at the clouds tops (Sromovsky et al., 2013; section 13.4).

The evolution from the localized eruption to the mature stage that constitutes the planetary-scale disturbance is dictated essentially by the structure of the ambient zonal wind at the outbreak latitude (Figure 13.3). The GWS evolution has been studied with some detail for the 1990 case using ground-based telescopes and the Hubble Space Telescope (Sanchez-Lavega et al., 1991; Beebe et al., 1992; Barnet et al., 1992) and in great detail for the 2010 case using a variety of ground-based telescopes and instruments onboard the Cassini spacecraft (Fletcher et al., 2011; Fisher et al., 2011; Sanchez-Lavega et al., 2011, 2012; Sayanagi et al., 2013; Sromovsky et al, 2014; section 13.3). These studies show that the latitudinal band occupied by the disturbance (Table 13.1) depends on the meridional size reached by the main spot during its growing stage and on the zonal wind meridional profile. The propagation of the disturbance along the latitude circles depends on the zonal wind velocity at each latitude but the details of the disturbance's cloud morphology depends, according to the models, on the thermal structure, vertical wind shear and Coriolis force (see section 13.7). For example, the mid-latitude 2010 event took place close to the peak on a westward jet and the disturbance propagated eastward forming a tail that encircled the planet in about 2 months (Sanchez-Lavega et al., 2012; Sayanagi et al., 2013). The 1990 Equatorial event took place in a sheared region of the strong eastward equatorial jet, and the disturbance propagated eastward (northern branch) and westward (southern branch) from the source, covering the whole latitude band in just one month (Sanchez-Lavega, 1994).

**[FIGURE 13.3]**

The convective head of the 2010-2011 GWS persisted for 7 months, until it was destroyed when the large anticyclone vortex AV of the disturbance arrived at its position; (Sanchez-Lavega et al., 2012; Sayanagi et al., 2013, García-Melendo et al., 2013). The underlying reason for the durability of the convective source for such a long period of time is still a mystery, although such persistence has also been observed in Jupiter's similar NTBD events (North Temperate Belt Disturbance) (Sanchez-Lavega et al., 2008).

The 1990 GWS was a huge event that disturbed the entire Equatorial Region with extended spot activity into mid-1991 (Sánchez-Lavega et al., 1993). Moreover, it re-activated and formed a large spot at 9.4°N that reached a size of ~ 27,000 km (zonal length) during 1994-95 (Sanchez-Lavega et al., 1996b). This activity coupled to vertical wind shear seems to be responsible for the change in the velocity of the equatorial jet between 1996 and 2002 (Sanchez-Lavega et al., 2003, 2004). The 2010 disturbance also produced a change in the profile of the westward jet where the storm emerged (Sayanagi et al., 2013; see section 13.3). These changes are consistent with adjustment to maintain thermal wind balance following a heating event at the latitude of the storm. The mechanisms responsible for this adjustment are not fully identified although shallow layer models point toward wave activity and momentum injection and dissipation (see section 13.7).

## 13.3. The 2010-11 Great Storm in the Cassini Epoch

The Great Storm of 2010-2011 provided the first-ever (and possibly once in a lifetime) opportunity to observe a GWS phenomenon from the vantage point of a spacecraft in orbit around Saturn.  Intervals of 27-30 years occurred between the five previous GWS events, with the last event happening in 1990; so the next event would have been expected in 2017-2020. The early occurrence of the 2010-2011 storm and the availability of the Cassini spacecraft was quite fortunate, as the planned end of the mission, when the spacecraft plunges into the atmosphere and is vaporized, is in September, 2017.  It remains to be seen whether the "2020 storm" will still occur, or the 2010-2011 storm disrupted the apparent 30-year cycle of the GWS storms.

In this section, we review the storm's temporal evolution, horizontal structure including the cloud morphology, and wind structure using the data captured by multiple instruments onboard Cassini as well as Earth-based observations. Among the Cassini instruments, the Imaging Science Subsystem (ISS; Porco et al 2004) returned high-resolution cloud images (Fischer et al 2011; Garcia-Melendo et al. 2013; Sayanagi et al. 2013, 2014; and Dyudina et al. 2013); the Visible and Infrared Mapping Spectrometer (VIMS; Brown et al. 2004) returned spatially resolved spectra that allow mapping of cloud composition (Sromovsky et al. 2013; Baines et al, 2015); the Composite Infrared Spectrometer (CIRS; Flasar et al. 2004) gathered stratospheric thermal structure and trace gas concentration (Fletcher et al. 2012, Hesman et al. 2012, Hurley et al. 2012), and the radar instrument (in the passive radiometer mode; Elachi et al. 2004) mapped the distribution of ammonia vapor (Janssen et al. 2013, Laraia et al. 2013). The Radio and Plasma Wave Science (RPWS; Gurnett et al. 2004) instrument can constrain the longitudes of lightning discharges through the timing of radio signals (Fischer et al. 2011, Sayanagi et al. 2013). Key observations were also made from Earth. In particular, ground-based observations using relatively small telescopes (many of which were operated by amateur astronomers) captured the development of the large scale disturbance (Sanchez-Lavega et al. 2011, 2012, Sanz-Raquena et al. 2012, cf, Chapter 14).

### *13.3.1 String of Pearls: A Precursor to the 2010-2011 Storm?*

The 2010-2011 storm erupted within an unusual feature known as the String of Pearls (SoP), detected in VIMS and ISS images (VIMS: Choi et al. 2009, Baines et al. 2010, 2015; ISS: Sayanagi et al. 2014). Figure 13.4 presents images of the SoP captured by the ISS (Sayanagi et al. 2014) and VIMS (Baines et al. 2015). In VIMS 5-micron images, the SoP appeared as a string of bright spots spanning some 90 degrees of longitude, which were interpreted to be holes in the cloud deck from which thermal radiation escapes from depth. In the ISS images, the SoP appeared as dark holes, which is consistent with the VIMS view. In the ISS images, the SoP were best viewed in the CB2 continuum filter (750-nm wavelength), and it was also visible in the visible bands (RGB) and CB3 filter (950 nm) albeit with lower contrast; they are completely indecipherable in MT2 and MT3, suggesting that the structure does not extend above the tropopause. Sayanagi et al. (2014) showed that the individual *pearls* have cyclonic vorticity, and the feature's longitudinal drift can be fitted by a constant speed of -22.27±0.2 m s$^{-1}$ (i.e., -2.26° of longitude per Earth day in System III; negative denotes westward) between September 2006 and July 2010. On December 5, 2010, the storm erupted at 114.1°E longitude and 32.6 °N planetocentric latitude (see Sánchez-Lavega 2011, for latitude definitions on an oblate planet). On December 5, the SoPs occupied longitudes between 62°E - 161°E. Thus, the storm's initial latitude-longitude coordinates coincided with those of the SoPs. The SoPs was last detected in an ISS image on December 23, 2010 (Sayanagi et al. 2014) before evidently being covered by higher altitude storm clouds or being physically disrupted by dynamics associated with the storm. By June 2011 the storm engulfed the entire latitude zone which subsequently became remarkably clear of clouds for several years through this time of this writing (2014), as indicated by the remarkably intense prolonged 5-μm brightness of the region that still exceeds any other region of the planet (Momary and Baines, 2014). With such a clear, cloud-free atmosphere, it is not surprising that the SoPs has not been seen since the storm.

**[FIGURE 13.4]**

### 13.3.2 Temporal Evolution of the 2010-2011 Storm

The storm erupted on December 5, 2010, as serendipitously detected in an ISS image (Sayanagi et al. 2013) and observed by ground-based amateur astronomers (Sanchez-Lavega et al. 2011). The Cassini RPWS instrument also detected an onset of electrostatic bursts that would continue until August 2011 (Fischer et al. 2011a, Sayanagi et al. 2013). On December 5, the storm spanned 1920±130 km east-west and 1300±130 km north-south, covering an area of approximately $1.5 \times 10^6$ km. Outside the storm, the surroundings appeared undisturbed from the relatively placid appearance it had maintained since emerging from the ring shadows around 2006 (Sayanagi et al. 2013).

The temporal evolution of the cloud morphology in the storm region is illustrated in the Cassini/ISS imagery shown in Figure 13.5. The red, green and blue color channels of the mosaic are assigned, respectively, to images acquired in a continuum band (CB2), a moderate methane absorption band (MT2), and a strong methane absorption band (MT3). Because methane is well-mixed, features that appear bright in the MT2 and MT3 absorption bands must reside at relatively high altitudes (Tomasko and Doose, 1984). Consequently, in this color scheme, the altitude of detected features generally increases in order of red-green-blue--white indicates that light is scattered in all three wavelengths bands by high-altitude aerosols.

**[FIGURE 13.5]**

At the westernmost point of the storm (Figure 13.5), there is a bright cloud identified as the storm's convective *storm head*, where, from VIMS spectra, Sromovsky et al (2013) detected water and ammonia ice lofted over 50-250 km in altitude, the first detection of water ice ever observed on Saturn.

Bright clouds billow to the east of the head out to the large anticyclonic vortex (AV). To the east of the AV, clouds acquire a turbulent appearance without well-defined edges, identified as the storm's *tail*. The storm's cumulus convection appeared to be confined to the area between the head and AV. Here, the most reflective storm clouds appeared (Sanchez-Lavega et al. 2012, Sayanagi et al. 2013), within which lightning flashes were observed in ISS imagery (Dyudina et al. 2013), and electrostatic radio bursts were detected by RPWS (Fischer et al. 2011a, Sayanagi et al. 2013). This region is identified as the *body* of the storm.

The AV formed between December 5 and 24, 2010. Initially, its core was centered at 76°E longitude and 33°N planetocentric latitude and appeared roughly circular with a diameter of approximately 4250 km. It appeared particularly bright in the MT3 strong methane absorption band (blue in Figures 13.5 and 13.6), indicating an unusually high altitude, as verified in a detailed multi-wavelength analysis of ISS images in CB2, MT2 and MT3 filters( Sayanagi et al. 2013). By January 11, 2011, the AV measured 11,000 km by 12,000 km in the north-south and east-west dimensions, a size that rivaled Jupiter's Oval BA (Simon-Miller et al. 2006).

Between the onset of the storm on December 5, 2010 and the head's last sighting on June 14, 2011, the *body* of the storm between the head and AV grew longitudinally due to the difference in their drift speeds, with the storm's head drifting westward at 2.79±0.08° of longitude per Earth day (26.9±0.8 m s$^{-1}$) (Sanchez-Lavega et al. 2011, 2012; Sayanagi et al. 2013) vs 0.85° longitude per day (8.4 m s$^{-1}$) for AV. As the space between the head and the AV widened at ~ 2.0° per day, the area covered by the storm's body changed from $1.75 \times 10^8$ km$^2$ on December 24, 2010, to $2.82 \times 10^8$ km$^2$ on January 2, 2011, and to $4.62 \times 10^8$ km$^2$ on January 11, 2011 (Sayanagi et al. 2013), corresponding to an approximately constant expansion rate of 140 km$^2$ s$^{-1}$. Sanchez-Lavega et al (2011) measured a peak expansion rate of 212 km$^2$ s$^{-1}$ at the cloud top height of 150 mbar during December 14-16. As the body of the storm grew, its boundary developed wave-like undulating patterns (Sanchez-Lavega et al. 2011, 2012). The Cassini RADAR radiometer examined this region and found that the body of the storm appears unusually bright in 2.2-cm wavelength thermal emission, indicating, rather surprisingly, that the region is sub-saturated in ammonia vapor (Janssen et al. 2013, Laraia et al. 2013). Li and Ingersoll (2015) show, in an axisymmetric numerical model, that this comes about due to uplift and precipitation of ammonia followed by a dry downdraft during the geostrophic adjustment process that follows a moist convective event. Dyudina et al. (2013) revealed that lighting flashes occurred in the body of the storm away from the head, suggesting that moist convection is occurring in the body as well. One potential explanation is that the thick clouds obscured lightning in the most convective parts of the clouds from view In late June 2011, six months after its formation, the head of the storm caught up with the AV after the vortex's position fell 360° of longitude behind (to the east) of the head due to the AV's slower westward drift (Sanchez-Lavega et al. 2012, Sayanagi et al 2013). Figure 13.5 panel 2011-06-14 shows the head and the AV 10 days before their positions were predicted to intersect on June 24. Ground-based observation by Sanchez-Lavega et al. (2012) showed that the head and the AV made contact around June 15, and the bright cloud materials of the head first disintegrated into three parts before disappearing between June 14 and June 19. On July 12 (Figure 13.5 panel 2011-07-12), the Cassini ISS imaged the storm for the first time after the predicted collision between the head and the AV; after the collision, the head was no longer visible while the AV was still present.

After the head-AV collision and the ensuing end of the convective activity, a dark featureless region appeared in the storm-affected planetocentric latitudes between 31°N and 38°N (Sayanagi et al. 2013). The region appeared dark in CB2 and visible continuum wavelengths and had a relatively bright appearance in MT2 and MT3, consistent with the absence of clouds (Sayanagi et al. 2013). Momary and Baines (2014) showed that the region is intensely bright in 5-μm thermal emission of light escaping from the deep atmosphere; indeed, it is the brightest 5-μm region observed thus far on Saturn, thus indicating a clear, almost cloudless "desert" of air in the troposphere down to several bars, consistent with the ~ 2--bar-level desiccation of ammonia vapor observed by the Cassini radiometer (Janssen et al 2013; Laraia et al., 2013). Thus what was perhaps one of the most powerful storms ever witnessed anywhere in the Solar System left in its wake a dry, cloudless, almost barren sky that has now lasted for several years.

### 13.3.3 The Effects of the Storm on the Wind Field

The storm had a significant impact on the local wind field as well as the zonal mean wind speed. Dyudina et al. (2013) and Garcia-Melendo et al (2013) found that the head of the storm had a large-scale anticyclonic circulation. In addition, Garcia-Melendo et al (2013) found a diverging motion in the head, suggesting that the cloud-top was an anvil cloud spreading out of a cumulus convective core. Dyudina et al. (2013) also showed each crest in the undulations on the north side of the storm body is associated with an anticyclonic vortex; the large anticyclone that defined the eastern edge of the storm body was the first of the series of anticyclones that line the storm every 10-15° in longitude.

Garcia-Melendo et al (2013) showed that peak wind speeds in the anticyclonic circulation around the head of the storm reached 160 m s$^{-1}$ with a relative vorticity reaching -9×10$^{-5}$ s$^{-1}$. Dyudina et al. (2013) reported that each of the anticyclones in the body of the storm had a peak speed reaching 100 m s$^{-1}$ and a relative vorticity of -5×10$^{-5}$ s$^{-1}$. Sayanagi et al (2013) also measured the wind structure of the easternmost anticyclone and showed that the peak wind speed exceeded 120 m s$^{-1}$, and the relative vorticity reached -6×10$^{-5}$ s$^{-1}$.

Sayanagi et al (2013) showed that the zonal mean wind speeds in the region affected by the storm were different before and after the storm (Figure 13.6). They showed that the zonal wind profiles before the storm (on May 7, 2008, and September 7, 2010) and in the early phase of the storm (on January 11, 2011) were essentially the same. After the end of the convective phase (on August~5, 2011), the wind speed changed. The planetocentric latitudes to the south of 34°N exhibit a deceleration of the zonal wind profile by ~30 m s$^{-1}$ while the profile to the north indicate an increase by ~35 m s$^{-1}$. Sayanagi et al. (2013) pointed out that these changes are consistent with warming at the central latitude of the storm and thermal wind balance to the north and south. Interestingly, the zonal wind measurements during the Voyager flybys in 1980-81 by Sanchez-Lavega et al (2000) indicate wind speeds between these pre- and post-storm measurements.

### [FIGURE 13.6]

Together with the thermal wind equation, Fletcher et al (2011) used Cassini CIRS temperature measurements to estimate the change in the vertical shear between September 2010 and January 2011 - that is, just prior to and after the advent of the storm. They detected wind speed changes greater than 40 m s$^{-1}$ above the 50-mbar level. In the deeper 1-bar to 100-mbar region, Fletcher et al. detected little change, consistent with the Sayanagi (2013) cloud-track measurements noted above for early storm measurements at comparable altitudes. Later in the storm development, Achterberg et al (2014) and Fletcher et al (2012) presented CIRS data that showed that, at the storm latitude, the tropospheric temperature increased between January and August of 2011. Achterberg et al. measurements indicate d$T$/d$y$ > 0 equatorward of the storm and d$T$/d$y$ <0 poleward of the storm (with $T$ being temperature and $y$ being the northward distance). The temperature gradient suggests that the wind speed above the 1-bar level decreased to the south of the storm and increased to the north. Thus, the CIRS vertical shear measurements are consistent with the interpretation by Sayanagi et al (2013) that the change in the cloud-tracked winds post July 2011 represents a real wind speed change.

## 13.4. Vertical cloud structure and composition

The Great Storm of 2010-2011 was the most well-studied meteorological event ever observed on Saturn, enabling detailed state-of the-art analyses and modeling of cloud, temperature, compositional, and dynamic structure. This event was truly four-dimensional in nature, with its morphology and thermal characteristics changing widely over latitude, longitude, depth and time. Here, we first explore the vertical structure and composition of the aerosols and clouds ad constrained from visual and near-infrared images, maps, and spectra. The temperature structure and dynamics are covered in Sections 13.5 and 13.7, while results on lightning emissions are covered in Section 13.6.

### 13.4.1 Models of the cloud structure from visible wavelength imaging

As previously discussed, the storm was first detected in visible light using small-aperture, ground-based telescopes (Fischer et al. 2011a; Sánchez-Lavega et al., 2011) and profusely imaged with high resolution by the ISS camera onboard the Cassini spacecraft in subsequent months (Sánchez-Lavega et al., 2012; Sayanagi et al., 2013). Most of the dynamical simulations of the storm (e.g. García-Melendo et al., 2013) were therefore based on the visual aspect of the storm and the evolution of the cloud field as imaged in wavelengths not far from those visible by the human eye. For this reason, even though longer wavelengths are more sensitive at higher and lower atmospheric levels and provide complete information on the distribution and nature of particles inside the storm (e.g., Sromovsky et al., 2014),), some information is also required on how the visible light is scattered, mostly by the uppermost particles raised by the convective event. This provides altitude levels for the location of cloud tracers and thus the cloud-tracked winds and possible wind shears, which can then investigated and compared against dynamical simulations. As well, historically storms have been viewed almost exclusively in the visible; therefore, any comprehensive comparison of the vertical cloud structure between past storms involves the analysis of images and spectra taken at wavelengths below 1 μm.

The initial report on the Great Storm of 2010-2011 by Sánchez-Lavega et al. (2011) included the estimation of the cloud top pressure and other parameters of the area disturbed by the storm. This work used images taken at a variety of filters including wide- (U, B, V, R) and narrow-band filters (624, 727, 883, 889, 914 and 954 nm), all of the latter related to red and near infrared methane absorption bands. The center-to-limb variation of the absolute reflectivity showed the GWS to be embedded inside the tropospheric haze. The main cloud deck was below the levels sounded by some filters, such as the deep methane band at 889 nm, which showed the disturbed area as darker than its surrounding. This implied that the haze above the storm itself had largely dissipated in some way. With such a haze-free atmosphere only the storm clouds contributed to the observed reflectivity, which showed the storm particulates to be substantially more reflective at all wavelengths, including the normally absorbing blue wavelengths. This suggested for the first time that the material comprising the main cloud deck in the disturbed area was fresh, less blue-absorbing than the common hazes and clouds found in Saturn (Pérez-Hoyos et al., 2005; Karkochka and Tomasko, 2005; West et al., 2009).

A similar approach was taken by Sanz-Requena et al. (2012) with observations taken one month after the outbreak of the storm. Wavelength coverage was from near ultraviolet to the near infrared including methane bands of various strengths. As in Sánchez-Lavega et al. (2011), they found that the GWS cloud tops reached the altitude level $300\ ^{-100}_{+200}$ mbar higher than the ~ 1400-mbar ammonia condensation level. Similarly García-Melendo et al. (2013) performed radiative transfer models on high-resolution images taken in February 2011 by Cassini ISS instrument in blue (BL1) and near-infrared (MT2, CB2, MT3) filters including two methane bands and an intermediate continuum (Figure 13.7). They also found cloud-tops at 400 mbar. The second most important change produced by the GWS was an overall increase in the single scattering albedo of the particles, with values close to 1 even at blue wavelengths. These results were analyzed and discussed in the context of the dynamical modeling by García-Melendo et al. (2013) and found to be in agreement with their numerical models of the 2010 event.

**[FIGURE 13.7]**

In short, there are three main characteristics of the 2010 GWS that could be retrieved from imaging at wavelengths below 1 μm: (1) the GWS can be modeled as elevated thick clouds extending up to levels between 300-400 mbar; (2) the particles injected into the atmosphere are substantially more reflective at all wavelengths than those commonly found at the same levels in Saturn's atmosphere; and (3), the levels above the thick clouds are not substantially enriched in particles, so they appear darker than their surroundings in the 889 methane absorption band, except perhaps at the front of the disturbance. Such results are in relatively good agreement with the analysis of the 1990 GWS by Acarreta and Sánchez-Lavega (1999) based on the imaging by Sánchez-Lavega et al. (1994) and Barnet et al. (1992) which placed cloud tops at 300 mbar.

There is still much work to be done regarding the analysis of the visible data below 1 μm. Firstly, the evolution of the cloud field showed a complex evolution in the mature phase of the event (Sánchez-Lavega et al., 2012; Sayanagi et al., 2013), and there are currently no works describing the corresponding evolution of the vertical cloud structure above the 1-2 bar level. Secondly, the properties of the small particles potentially located above the storm could be best sounded using visual wavelengths. Indeed, the large coverage of phase angles provided by the Cassini spacecraft in 2011 can be used to explore the size or phase function of high-altitude particles. Finally, since the blue absorption of tropospheric particles remains as one of the most elusive problems in the atmospheres of Jupiter and Saturn, how the storm affected its color could also provide a substantial piece of information on this issue.

### 13.4.2. Composition of cloud particles from VIMS spectral images

13.4.2.1. VIMS observations of the storm

The VIMS instrument (Miller et al. 1996; Brown et al. 2004) covers a spectral range of 0.35-5.1 μm, with an effective pixel size of 0.5 milliradians on a side and near-IR spectral resolution of approximately 15 nm. Samples of the earliest VIMS observations of the Great Storm, acquired on February 24 2011are shown in Figure 13.8 as remapped equirectangular projections at four key near-IR wavelengths. The most dramatic spectral characteristic of the storm is its low reflectivity at 3.05 μm (panel B), something not previously seen anywhere else on Saturn (Sromovsky et al., 2014). Outside the storm region the *I/F* at 1.887 μm (panel A) is a good match to the *I/F* at 3.05 μm, but in the storm head the *I/F* at 3.05 μm is only about 40% of the outside value. At 4.081 μm, the storm is nearly twice as reflective as non-storm clouds, indicating the presence of larger particles in the storm region. At 5.041 μm (panel D), the thermal emission of Saturn's deeper layers (near 3 bars) contributes to the observed *I/F*. The dark regions of the storm head at this wavelength indicate the presence of optically thick cloud layers. The spectral characteristic of the 3-μm absorbing region is illustrated in Fig. 13.8E, where a spectrum from inside the storm head (solid line) is compared with a spectrum from the region just west of the storm head (dotted line). Absorption by $PH_3$ affects the 2.8-3.03 μm region, but the spectral difference seen in the 3.03-3.14 μm region is almost entirely due to particulate absorption. Spectra obtained upstream display absorption between 2.8 and 3 μm that is due to phosphine, but show little absorption at 3.05 μm. The upstream clouds are also more transparent at thermal wavelengths, evident from their large apparent *I/F* values near 5 μm. The storm particles are brighter than the upstream cloud particles in most window regions, slightly more absorbing at 2.7 µm, and dramatically more absorbing at 3.048 µm.

**[FIGURE 13.8]**

13.4.2.2 Vertical structure models

Near-IR spectra outside the storm regions can be very well fit by relatively simple cloud structures similar to those used successfully to fit spectra in the CCD range from 0.3 to 1 µm (Muñoz et al. 2004; Karkoschka and Tomasko 2005; Pérez-Hoyos et al. 2005). These structures contain a vertically diffuse stratospheric haze and an optically thick and physically thick main cloud layer, which requires visible and UV absorption, but can fit near-IR spectra with conservative cloud particles. Sromovsky et al. (2013) used a similar structure to obtain the excellent fit shown by the model spectrum (gray curve) in Fig. 13.8E. For this fit the main haze layer contained Mie particles of real index 1.4 and effective radius 0.6 µm, extended from 120 mbar to 700 mbar, and had an optical depth of 6.8 at 2 µm. An optically thick layer with an effective pressure near 3 bars is also needed to control thermal emission.

13.4.2.3 Single-component models of 3-µm absorption

Sromovsky et al. (2013) first tried to fit the storm spectra using pure substances in a single middle cloud layer of spherical particles, then adjusted the optical depth, particle size, and top and bottom pressures to optimize fits for the region from 1.268 µm to 5.15 µm, while ignoring the region from 2.65 µm to 3.2 µm where particulate absorption is most prominent. That allowed the vertical structure of the aerosols to be primarily constrained by methane gas absorption and not much affected by particulate absorption. They then compared the model spectra with measured spectra in the 2.65 µm - 3.2 µm region to determine which provided the best compositional match. They considered five possible cloud materials: (1) a conservative substance with refractive index n=1.4+0i, (2) $NH_3$, (3) $NH_4SH$, (4) $N_2H_4$, and (5) water ice. The fits for the five pure substances are displayed in Figure 13.11. Ammonia provided by far the best match to the measured spectrum of the storm head, yielding a $\chi^2$ value in the absorbing region that is 3.5 times better (smaller) than that of the next best fit. However, this best spectral fit among these simplistic single-component models, still contains substantial differences from the VIMS measured spectrum, especially the sharp ammonia ice absorption feature near 2.96 µm, but also the excess brightness near 3.1 µm.

13.4.2.4 Multi-component models of 3-µm absorption

Sromovsky et al. (2013) next considered linear combinations of pure spectra, which can be thought of as spatially heterogeneous mixtures of different cloud types, perhaps as different particles visible in spaces between clouds of other particles. The best linear combination, shown by the blue curve in Figure 13.9B used a mix of 55±4% $NH_3$, 22±2% water ice and 23±2% $NH_4SH$, where percentages represent fractional area coverage. This improved the fit in the absorbing region to $\chi^2/N(ABS) = 1.1$, a factor of 8 better than provided by pure $NH_3$ (see Fig. 13.9 caption). A big improvement was also obtained using the conservative (n=1.4) material in place of $NH_4SH$, with a mix of 55±3% $NH_3$, 22±3% water ice, and 23±2% n=1.4, although the improvement was only by a factor of five in this case. The best linear combination without water ice produced $\chi^2/N(ABS) = 5.2$, which is a factor of 5 worse than produced by the best combination with water ice. The water ice component reduced the excess $I/F$ gradient between 3 µm and 3.15 µm and the excess $I/F$ otherwise present between 2.65 µm and 2.85 µm. The third component (either n=1.4 or $NH_4SH$) served to partially fill in the deep minimum produced by pure $NH_3$ at 2.96 $\mu$m. Their linear combination analysis clearly favors $NH_3$, water ice, and $NH_4SH$ in that order.

**[FIGURE 13.9]**

They also considered more physically plausible locally horizontally homogeneous cloud models, for which their best fit to VIMS spectra contained a deep water cloud, perhaps beginning at 10 bars, but certainly containing significant opacity above the 1-2 bar range, a main cloud layer consisting of two sub-layers, one of low optical depth ($\tau \approx 1$) above a lower sub layer of significant optical depth ($\tau \approx 9$). Among two options for the main layer that produced good fits, the most plausible was one with particles of n=1.4 in the top sub-layer and $NH_3$ in the lower sub-layer as a coating on $H_2O$ cores, with a core radius that is 40% of the total radius (1.5 $\mu$m), and a core volume of 6%. A slightly better fit was obtained using smaller particles with some long-wave absorption in the upper main cloud layer and somewhat larger $NH_3$-coated water ice particles in the lower main sub layer. Such a model is qualitatively consistent with deep convection from the water layer, as suggested by Hueso and Sánchez-Lavega (2004). Both heterogeneous and homogeneous models strongly favored $NH_3$ to be the most prominent substance producing 3-$\mu$m absorption, with water ice providing an important contribution in both cases.

The relative roles of $NH_4SH$ and the n=1.4 material were less clearly defined. It is also plausible that storm cloud particles might contain three materials: a core of water ice, a shell of $NH_4SH$, and a final coat of $NH_3$. Whether such particles would lead to better spectral fits remains to be determined. Also somewhat unclear is how much these relative contributions might be changed by revising the VIMS calibration (Clarke et al., 2012).

A firm conclusion of the VIMS Near-IR study of Sromovsky et al (2013) is that frozen water is indeed present in the convective storm head of the Great Storm of 2010-2011. The presence of water ice lofted 200 km above its freezing level near 10 bar proves that such Saturnian storms are much more powerful, and extend over much more extensive depths than their terrestrial counterparts.

## 13.5. Temperature Structure

Optical studies of the storm provided important insights into the physical processes at work within Saturn's cloud-forming regions, but the presence of Cassini in the Saturn system, combined with two decades of technology development in the mid-infrared, meant that Saturn's 2010 storm was the first to be studied in detail at thermal infrared wavelengths. Thermal infrared studies provide insights into the environmental conditions (temperature structure and gaseous composition) both within and surrounding the storm system. By reconstructing the temperature field in three dimensions, and monitoring its evolution through time, we gained powerful new insights into the effects of a tropospheric storm system on the overlying stratosphere, including the birth of an enormous hot anticyclonic vortex (referred to as the "*beacon*") high above the storm region. This enormous stratospheric vortex persisted long after the tropospheric convective activity had abated and was last observed in 2013, although observations in 2013-14 still showed residual wave activity in Saturn's stratosphere. Thermal contrasts driven by the storm were first detected by Fletcher et al. (2011), and their temporal evolution has been reported in both the troposphere (Achterberg et al., 2014) and the stratosphere (Fletcher et al., 2012; Hesman et al., 2012) using a synthesis of data from the Cassini Composite \Infrared Spectrometer (CIRS) and a host of ground-based observatories (notably the Very Large Telescope, VLT, and Infrared Telescope Facility). The stratospheric response to the tropospheric storm has resulted in the discovery of the greatest atmospheric temperature change ever seen due to a storm (Fletcher et al. 2012; Hesman et al. 2012) in addition to completely unexpected changes in species abundances (Hesman et al. 2012, Bjoraker et al. 2012, Hesman et al. 2013). This section summarizes what has been discovered about Saturnian storms (and therefore giant planet convective plumes in general) from assessments of the thermal structure.

### 13.5.1 Tropospheric Temperatures

Remote sensing in the mid- and far-infrared (5-1000 μm) is typically sensitive to altitudes above the topmost condensate clouds but below the tropopause (approximately 80-700 mbar), in addition to sensitivity in the mid-stratosphere (0.1-10 mbar from nadir observations) from hydrocarbon emissions (primarily methane, ethane and acetylene). Fletcher et al. (2011) presented the first 7-25 μm images of the storm region on January19, 2011 using the VLT/VISIR instrument, 45 days after the initial eruption of cloud material. Troposphere-sensitive images (Fig.13.10) were centred on the cold core of a newly-formed anticyclonic vortex embedded in the westward jet near 39°N (planetographic), which appeared dark bluish in amateur imaging surrounded by the collar of whiter cloud material. The cold oval near 41°N had a core temperature of $79 \pm 2$ K at 100 mbar consistent with upwelling and adiabatic cooling within an anticyclone of approximate dimensions of 4000 x 5500 km. To the west of the oval, the bright white storm head extended 60° longitude away from the anticyclone, and was characterized by warmer tropospheric temperatures ($5 \pm 2$ K warmer than the atmosphere east of the storm anticyclone in the 70-300 mbar region) surrounding discrete cold spots. The warmest regions appeared restricted to the southern edge (35-38$^{o}$N) of the storm band, possibly co-located with the southern band of bright white clouds.

**[FIGURE 13.10]**

To the east of the anticyclone, the VLT images in Fig. 13.10 show that the flow bifurcated into two cool branches (Fletcher et al., 2011): a northern branch at 42° - 48°N and a southern branch between 31 - 38°N separated by a fainter warm sector. These two branches were characterized by undulating white clouds and contrasts between warm and cooler spots, suggestive of atmospheric thermal waves remaining in the wake of the westward-moving storm head. Although VLT images provided only a snapshot of the thermal field on a few days during the storm's lifetime, they were capable of resolving the small-scale longitudinal structures (i.e., small plumes) that would be hidden to Cassini's spectrometer (CIRS) because the signal-to-noise level on individual CIRS spectra precludes the generation of high-resolution tropospheric maps. Nevertheless, Cassini/CIRS spectra are essential to provide the spectral information necessary to retrieve the vertical temperature structure.

Achterberg et al. (2014) utilized 20-200 μm spectra of Saturn's smooth $H_2$-He collision-induced continuum to determine zonal mean temperatures and para-$H_2$ distribution, monitoring these parameters from 2009 to 2012 to understand the tropospheric evolution. During the storm, they discovered that the zonal mean temperatures for p > 300 mbar increased by 3 K, accompanied by a fall in the para-$H_2$ fraction that is consistent with the idea of upwelling associated with the convective storm. As Saturn's winds are in approximate geostrophic balance, Achterberg et al. (2014) use the rising temperatures to show that the post-storm winds become more eastward with height poleward of the storm band, and more westward with height equatorward of the storm band, by 5 m/s per scale height. This is in approximate agreement with the wind speeds determined by Sayanagi et al. (2013) from cloud tracer measurements.

No complete picture of the storm-induced tropospheric heating has yet emerged. Vertical advection of air heated by the latent heat from water condensation could have warmed the storm head during the active phase (Fischer et al., 2011a, Achterberg et al., 2014), particularly if restricted to the southern edge of the storm band as shown in Fig. 10B. Alternatively, a thermally-indirect circulation forced by waves and eddies could result in localized updrafts cooling due to adiabatic expansion in the stably-stratified upper troposphere (the colder spots in Fig. 10B), surrounded by a larger area of subsidence and adiabatic heating by compression. Indeed, microwave observations in the post-storm phase suggest $NH_3$ depletion consistent with atmospheric subsidence over wide areas of the storm remnant (Janssen et al., 2013; Laraia et al., 2013). Such small-scale uplift, balanced by subsidence over a wider area, is a typical hallmark of a moist-convective process, and we note that plumes associated with the revival of Jupiter's South Equatorial Belt also tend to be cold-cored (Giles et al., 2013) and followed by large-scale warming. It remains unclear whether the rise in temperature in the storm band was due to moist convective heating on an updraft or adiabatic warming on a downdraft; whether it flipped from the former to the latter at some stage during the storm's evolution; and whether warming by latent heat might dominate the deep troposphere whereas warming by large-scale subsidence dominates the upper troposphere. It was shown that the storm increased the emitted power in a wide latitudinal band (20–55°N) with a maximum change of 9.2 ± 0.1% around 45°N from 2010 to 2011 (Li et al., 2015). Tropospheric temperatures appeared to still be rising in early 2012 (Achterberg et al., 2014) and IRTF imaging in 2013 (Fletcher et al., 2014) showed the continued presence of the warm tropospheric band near 40°N, consistent with subsidence in the band causing desiccation and depleting the atmosphere of aerosols remaining from the convective storm (the band is extremely bright when viewed at 5 μm, Orton et al., 2013).

### 13.5.2 Stratospheric Temperatures

Perhaps the greatest surprise revealed by the thermal infrared studies was the intense perturbation created in Saturn's stratosphere, completely invisible to observers using reflected sunlight. Emissions from stratospheric hydrocarbons (namely methane, ethane and acetylene) permit the retrieval of the stratospheric thermal structure, under the assumption that methane remains well-mixed despite the powerful dynamics associated with the storm. North-south and east-west retrievals of 0.5-5 mbar temperatures from Cassini suggested a substantial stratospheric cooling over a wide area approximately co-located with the tropospheric anticyclone (Fletcher et al., 2011). To the east and west, this cool region was flanked by two warm stratospheric `beacons' near 40°N, so-called because of their intensity when viewed in the infrared, and some 16 ± 3 K warmer than the cold centre in January 2011. As we shall describe below, the evolution and merger of these stratospheric beacons into a single large anticyclone dominated the appearance of Saturn's northern stratosphere for some time after the tropospheric storm had ended.

The origin of the stratospheric *beacon* continues to be a source of debate. Although some degree of convective overshooting to the 60-mbar level might have been expected from moist convection models (e.g., Hueso and Sánchez-Lavega, 2004), vertical mass motion from the storm genesis region should not penetrate high into the stratosphere. Saturn's atmosphere is so stably stratified in the tropopause region that updrafts would quickly become negatively buoyant, preventing further penetration into the stratosphere. Furthermore, even if the updrafts could be maintained by some mechanical forcing, the adiabatic cooling from expansion would be enormous, yielding regions at 1-mbar that were tens of Kelvin cooler than their surroundings, which was not observed. Thus wave propagation, generated by the strong mechanical forcing associated with the tropospheric disturbance, is a plausible explanation for the origins of the warm beacons. However, this is only part of the story – we initially see diffuse anticyclonic vortices instead of periodic temperature structures that we might expect from vertically propagating waves, suggesting a markedly nonlinear phenomenon. Furthermore, we might have expected these two initial beacons to cool and fade from view, but the subsequent evolution was completely unprecedented.

Fletcher et al. (2012) studied the evolution of the two warm regions from January 2011 to March 2012, dividing the evolutionary period into three phases as shown in Fig.13.11. This figure has been extended to early 2013, identifying a fourth phase in the vortex evolution where it was moving with a higher velocity. *Phase 1* (January to May 2011) saw the formation and intensification of the two warm regions (B1 and B2) with peak temperatures near 0.5 mbar. Both moved westward with time, with B1 residing directly above the tropospheric storm head and moving with the same velocity (2.73 ± 0.1°/day), and B2 moving more slowly but in the same direction (0.6 ± 0.1°/day). Over time, the storm head beacon caught up with the second, smaller warm region. These two warm air masses behaved like coherent vortices and merged when they encountered one another in late April 2011, forming a single large vortex (B0) with peak temperatures of 221.6 ± 1.4 K at 2 mbar (80 K warmer than the quiescent stratosphere) measured on May 5th 2011, immediately after the merger at the start of *Phase 2* (May to July 2011). This single vortex became disassociated from the tropospheric storm head and moved with an independent (slower) westward velocity (1.6±0.2°/day) that was intermediate between the initial velocities of B1 and B2. Furthermore, the peak temperatures in the vertical profile were found to be 1.4 scale heights deeper (approximately 162 km) post-merger in Phase 2, with the sudden subsidence presumably contributing to the rapid rise in temperature within the vortex. At its largest, B0 was 65° in longitudinal extent (50,000 km at 39°) and extended from 20 – 60°N latitude. In the final decay phases (*Phase 3 and 4*) after July 2011 (i.e., after the tropospheric storm had abated) were characterized by slow cooling (0.11 ± 0.01 K/day) and longitudinal shrinkage. During this phase, B0 accelerated from a constant velocity of 2.40 ± 0.29° /day at the end of 2011 (-25.0±3.0 m/s at 40°N during *Phase 3*) to 3.01±0.05°day throughout 2012/2013 (-31.3±0.5 m/s during *Phase 4*), meaning that it completed a circuit of the planet once every 120 days and was actually moving faster than the underlying retrograde jet in the troposphere. Guerlet et al. (2014) demonstrated that processes in addition to radiative cooling must be causing the temperature decay within the vortex, and that pure radiative decay would theoretically imply that the warm anomaly could persist until 2016. However, at the observed rate of cooling and shrinkage, Fletcher et al. (2012) predicted that the temperatures would return to their quiescent values by January 2013. CIRS observations have since suggested that the distinct vortex was absent in late 2013, having been replaced by substantial longitudinal wave activity that continues to create thermal contrasts within the 25 – 50°N region of the stratosphere as of the time of writing.

**[FIGURE 13.11]**

The 80-100 K stratospheric thermal contrast measured during *Phase 2* was unprecedented, and led to unique chemical and dynamical consequences within the vortex itself. In the vertical direction, temperatures appeared to drop towards the quiescent stratospheric temperatures for P < 2 mbar, although CIRS nadir observations lack sensitivity to P < 0.5 mbar and limb observations were not able to sample the vortex itself. Calculations of the static stability at lower pressures (above the peak temperatures at 2 mbar) suggest that the atmosphere remained sub-adiabatic and therefore stable to convection. Assuming no horizontal motion at the tropopause, and that the zonal and meridional temperature contrasts are in approximate geostrophic balance with the winds, integration of the thermal wind shear relation predicts that the hot airmass would be isolated by a peripheral clockwise wind of 200-400 m/s at 2 mbar (Fig. 13.12). As upwelling and divergent tropospheric plumes tend to inject anticyclonic vorticity, this suggested that B0 was a large stratospheric anticyclone only visible in the thermal infrared (Fletcher et al., 2012). Although the origins of the beacons was likely related to non-linear phenomena associated with waves rising from the tropospheric storms, the mature anticyclone can be likened to terrestrial anticyclones, which are regions of high pressure associated with subsiding, dry air. Extending this analogy to Saturn, friction at the lower boundary of the beacon (i.e., the tropopause) could cause outflow that is balanced by subsidence within the anticyclone to generate the warm temperatures by adiabatic compression. Indeed, after the merger the warmest beacon temperatures were at a higher pressure (2 mbar) and the beacon became visible at wavelengths of 17-20 μm, suggesting subsidence and a warming of the tropopause region.

**[FIGURE 13.12]**

Stratospheric anticyclones are common features of the Earth's stratosphere, most apparent in northern winter and southern spring (Harvey and Hitchman, 1996; Harvey et al., 2004), and possibly associated with the upward extension of planetary waves from tropospheric sources providing the source of energy (e.g., Charney and Drazin, 1961). On Earth, these stratospheric anticyclones are regions of strong gradients of potential vorticity and trace species (e.g., Manney et al., 1995). On Saturn, the peripheral collar apparently isolated a region of air enhanced in stratospheric hydrocarbons (Fig. 13.12), most notably acetylene ($C_2H_2$) and ethylene ($C_2H_4$), and a smaller enhancement in ethane ($C_2H_6$) (Fletcher et al., 2012; Hesman et al., 2012). These enhancements are most likely the result of the downward shift from 0.5 to 2 mbar (1.4 scale heights) post-merger, as stratospheric diffusive photochemistry models (Moses et al., 2015; Cavalié et al., 2015) have so far been unable to reproduce the chemical enhancements observed by Cassini without invoking downwelling winds. There was no evidence for effects of the stratospheric vortex in Saturn's tropospheric clouds and hazes as measured by Cassini/ISS and Cassini/VIMS (Fletcher et al., 2012). However, the evaporative effect on stratospheric hazes and water ice (condensed from externally-supplied oxygen species) has not yet been investigated, but these may provide an additional gaseous source for species found enhanced within B0. Finally, B0 left in its wake a new warm stratospheric band centred at 48-50ºN and 0.3-0.8 mbar. Thermal wind balance therefore suggests increasingly eastward flow with altitude at latitudes poleward of 50ºN and increasingly westward flow with altitude equatorward of 50ºN.

Although B0 has now met its demise, mysteries surrounding its origin remain unresolved. Convective plumes impinging on the tropopause act as strong sources of waves (gravity waves, planetary waves) which propagate in 3D away from the source. The stratospheric transmissivity to these waves depends on the details of the temperature and wind structure near the tropopause (and therefore the refractive index), and simple calculations by Fletcher et al. (2012) suggest that upward propagation of Rossby waves was permitted in the springtime epoch of this storm. Nonlinear effects and wave breaking may have led to the formation of the original vortices near 0.5 mbar, although this phenomenon has not yet been reproduced by any numerical simulation of Saturn's stratosphere. Indeed, the formation and evolution of Saturn's stratospheric beacon should be a key challenge for the next generation of coupled radiative-dynamical-chemistry models. Key observational constraints would come from the presence (or absence) of stratospheric vortices spawned by similar convective storms on Saturn and the other giant planets.

### 13.5.3. Stratospheric molecular composition

In this section reviewing stratospheric molecular composition, we continue to refer to the four phases of evolution of Saturn's stratospheric anticyclone: *Phase 1* when the two initial beacons were present and growing; *Phase 2* when the beacons had merged; and *Phases 3 and 4* during the decay phase of the mature beacon. The high stratospheric temperatures led to enhanced emission from both major hydrocarbon species (ethane $C_2H_6$ and acetylene $C_2H_2$, Fletcher et al., 2012; Hesman et al., 2012) and minor hydrocarbon species ($C_2H_4$, $C_3H_4$, $C_3H_8$, $C_4H_2$, Bjoraker et al., 2012). For many of the hydrocarbons, the enhanced emission cannot be solely attributed to the rise in atmospheric temperature (Bjoraker et al., 2012). Neither are they solely the result of the shifting photochemical reaction rates associated with the warmer temperatures (Cavalie et al., 2015; Moses et al., 2015). Indeed, the elevated abundances of the hydrocarbons can be used as diagnostic tracers of the subsiding motion within the stratospheric anticyclone, moving photochemically produced molecules downwards (e.g., Fletcher et al., 2012; Hesman et al., 2012; Moses et al., 2015).

The unique high temperatures, chemistry and dynamics of the stratospheric beacon resulted in the detection of ethylene ($C_2H_4$) by Cassini's Composite Infrared Spectrometer (CIRS) and the high-resolution ground-based spectrometer Celeste deployed on the McMath-Pierce telescope on Kitt Peak, AZ. CIRS was also used to identify other rare hydrocarbon species such as diacetylene ($C_4H_2$) and methylacetylene ($C_3H_4$) (storm *phases 1, 2, and 3*). Saturn's ethylene emission has been difficult to detect on Saturn (Encrenaz et al. 1975; Bezard et al. 2001). Indeed, prior to the 2010 storm, ethylene was not detected by CIRS. However, stellar occultation data acquired by Cassini's Ultraviolet Imaging Spectrograph (UVIS) yielded a vertical profile of $C_2H_4$ under pre-storm conditions between 0.1 μbar and 0.5 mbar at 15.2°N (Shemansky and Liu, 2012). This UVIS profile then provides a profile to compare the CIRS storm results at 0.5 mbar against. The May 2011 CIRS (Figure 13.15b) and Celeste (Figure 13.13a) spectra were used to determine a $C_2H_4$ mole fraction of $(5.9\pm4.5) \times 10^{-7}$ at 0.5 mbar (CIRS) and $(2.7\pm0.45)\times10^{-6}$ at 0.1 mbar (Celeste) (storm *phase 2*). These values are both two orders of magnitude higher than the amount measured by UVIS under pre-storm conditions. It is also much higher than predicted by photochemical models, indicating that perhaps another production mechanism is required or a loss mechanism is being inhibited (Hesman et al. 2012). Because ethylene leads directly to the formation of other more prominent hydrocarbons, such as ethane, understanding how it is created and destroyed is important to understanding the photochemistry in giant planet atmospheres. Full radiative transfer analyses are required to disentangle the effects of temperatures and abundances, and these have been performed for multiple hydrocarbon species by Hesman et al., 2012, 2013; Fletcher et al., 2012 and Moses et al., 2015. The results of from Hesman et al. (2013) are shown in Figure 13.13c and 13.13d. Scale factors were applied to profiles produced from photochemical models of Hesman et al., (2012). These results confirm that the $C_2$ hydrocarbons are significantly enhanced in Saturn's beacons compared with pre-storm values. $C_2H_4$ exhibits the greatest enhancement followed by $C_2H_2$. For the January 2012 dataset there is latitudinal structure within the beacon, with larger abundances for $C_2H_4$ at 33°N than at 38°N. And finally, the $C_3$ and $C_4$ hydrocarbon abundances exhibit only minor enhancement in Saturn's beacons, with emission that can be mostly explained by the rise in temperature.

**[FIGURE 13.13]**

Moses et al. (2015) and Cavalie et al. (2015) attempt to explain these hydrocarbon observations via photochemical modeling. Moses et al. find that the temperature rise would leave ethane, acetylene and propane relatively unaffected, whereas we would expect an ethylene increase due to the elevated temperatures. Nevertheless, these photochemical models still underpredict the observed abundances by factors of 2-7 in the beacon core in May 2011, shortly after merger. Downwelling winds of order 10 cm/s near 0.1 mbar, in addition to the aforementioned chemistry, are required to explain the elevated abundances in the $C_2$ hydrocarbons (Moses et al., 2015), although we caution that these 1D winds are likely a gross simplification of the 3D dynamics at work within this anticyclone. With these winds in place, the model shows that we would expect a factor of 2.2 increase in $C_4H_2$ (primarily driven by evaporation of ices), whereas the effects on $C_3H_8$ and $C_3H_4$ would be relatively minor, consistent with the preliminary findings of Hesman et al. (2013). Evaporation of water ice and $C_6H_6$ aerosols could also lead to increases in the gas phase abundances of water and benzene within the beacon.

In the coming years further analysis is needed to compare the hydrocarbon enhancements predicted from chemistry and dynamical models (including 3D simulations of the vortex dynamics) to the reality provided by Cassini and Earth-based studies of Saturn's hydrocarbon emission. Photochemical models should be updated with the time-dependent temperatures, species abundances and rate mechanisms. In addition, other sources of these species, such as aerosol vaporization, should be examined in conjunction with photochemical models. It is only by doing these studies that the compositional changes in Saturn's Great Storm of 2010-2011 will be understood.

## 13.6. Lightning events

The observation of Saturn lightning events has had a long history. The first indication of the existence of lightning in Saturn's atmosphere was obtained in November 1980 by the radio instrument on-board Voyager 1, which recorded strong and impulsive radio signals (Warwick et al., 1981), referred to as Saturn Electrostatic Discharges (SEDs), They were recorded again during the Voyager 2 flyby in August 1981, and, since 2004, by the RPWS (Radio and Plasma Wave Science) instrument (Gurnett et al., 2004) on-board the Cassini orbiter (Fischer et al., 2006). The multi-year presence of Cassini in Saturn's orbit has enabled the first observations of lightning during a GWS event (Fischer et al., 2011a) confirming – along with the water-laden cloud particle evidence mentioned in section 13.4.2.4 – that GWS events are giant thunderstorms with strong vertical convection (e.g. Sánchez-Lavega, 1994).

### 13.6.1 Properties of Saturn Electrostatic Discharges

The SEDs are short-duration (~100 ms) impulsive radio bursts with a large frequency bandwidth. In order to penetrate Saturn's ionosphere and reach the spacecraft their frequency has to be greater than the ionospheric cutoff frequency, the latter being of the order of a few hundred kHz to a few MHz on the nightside and dayside, respectively (Kaiser et al., 1984; Fischer et al., 2011b). In the frequency range of SEDs the RPWS instrument acts as a sweeping receiver that detects the radio burst at whatever frequency the receiver happens to be tuned to at the time of the flash. This leads to the typical occurrence of SEDs as short vertical bursts in the time-frequency spectrum of Fig. 13.14, despite the fact that short bursts as SEDs are intrinsically broadband emissions. Fig13.14 shows that single vertical bursts can only be discerned around the edges of the SED episode, whereas in its center and especially around 19:30 SCET the flash rate is so high that the entirety of SED bursts appear as a patchy emission with no isolated single bursts. Here the observed peak flash rate is >10 s$^{-1}$, a rate that surpasses the typical flash rates of smaller SED storms by about 2 orders of magnitude. Another remarkable feature of the SED episode in Fig. 13.16, recorded during the GWS on February 26, 2011, is its long duration of almost 8 hours (13:20 to 21:20 SCET). The frequency range of SEDs in Fig. 13.16 ranges from ~16 MHz (highest frequency of the RPWS instrument) to a lower cutoff frequency near 1-2 MHz. SEDs can be used as a natural tool to study Saturn's ionosphere The measurement of the lower cutoff frequency combined with the knowledge of the SED source position is the only method which can reveal peak electron densities of Saturn's ionosphere at all local times (Fischer et al., 2011b).

**[FIGURE 13.14]**

Before we discuss the SED properties observed during the 2010-2011 GWS in more detail, we highlight the main properties of regular (smaller) SED storms as they were observed from 2004 to 2010 by the Cassini spacecraft (blue dots in Fig. 13.1). The typical flash rate of regular SED storms observed by Cassini is a few SEDs per minute (Fischer et al., 2008). Most of the 10 regular SED storms observed from 2004 until 2010 lasted for several weeks to months, and the longest one raged in 2009 for almost the entire year. However, there were also long intervals of time with no SED activity, e.g. from February 2006 until November 2007. The SED storms were correlated to cloud features of about 2000 km in diameter located at the so-called storm alley at planetocentric latitude of 35°S (Fischer et al., 2011c; Baines et al, 2009). The first link between SEDs and optical storm cloud observations was found by combined observations of the Cassini cameras and the RPWS instrument (Porco et al., 2005). They revealed consistent longitudes and longitudinal drift rates of the cloud feature in comparison to the SED occurrence, and the white storm clouds were also found to be brighter when the SED rates were high (Dyudina et al., 2007; Fischer et al., 2007a). A white cloud captured by Cassini/VIMS in February, 2008 was also observed to be at the longitude where SED lightning was being contemporaneously observed (Baines et al., 2009). The prominent 2000-km sized storm clouds turned out to be easily detectable by amateur astronomers on Earth (Fischer et al., 2011c), who for almost a decade have provided valuable Saturn storm cloud observations on the website of the Planetary Virtual Observatory (Hueso et al., 2010).

SEDs usually occur in episodes lasting for about 6-7 hours, which is somewhat longer than half a Saturn rotation (~5 h 20 min or over 180° in sub-spacecraft longitude). This is due to the fact that most SED storms consist of a single storm cell from which SEDs are observed when the storm is on the side of the planet facing the spacecraft (the orbital period of Cassini is always much longer than a Saturn rotation). A radio wave propagation effect also allowed SEDs to be detected "over the horizon" by up to 45° in longitude when the SED source rose or set at the nightside horizon (Zarka et al., 2006). Consequently, an SED episode lasting for 8 hours (or over ~270° in sub-spacecraft longitude) as in Fig. 13.15 must be caused by at least two or several longitudinally separated storm cells.

**[FIGURE 13.15]**

SEDs have a spectral radio source power of the order of 10 to 100 W/Hz, which is about 10,000 times stronger than the radio signals of terrestrial lightning in the frequency range of a few MHz (Fischer et al., 2006; Zarka et al., 2006). Taking an SED duration of 0.1 s and an SED frequency range of 10 GHz (similar to terrestrial lightning) one would arrive at a value of $10^9$-$10^{10}$ J for the radio energy of one SED, which is similar to their optical energy (Dyudina et al., 2013). This is only a rough estimate since we have no information on the SED frequency range and do not know how the radio intensity varies with frequency. The high SED intensity (flux ~100 Jansky at Earth) and Cassini's reports of on-going lightning activity enabled the first SED detection by the large ground-based radio telescope UTR-2 in the Ukraine (Zakharenko et al., 2012; Konovalenko et al., 2013).

Below ~2 MHz SEDs are highly polarized emissions with a high degree of circular polarization. Fischer et al. (2007b) found SEDs to be right-hand polarized when emitted from the southern hemisphere, and they predicted that SEDs would be left-hand polarized when emitted from the northern hemisphere, a difference that is related to the direction of the magnetic field vector in the two hemispheres relative to the radio wave propagation vector. This prediction was confirmed by polarization measurements of SEDs from the 2010-2011 GWS, which was the first SED storm located in the northern hemisphere during the Cassini era (Fischer et al., 2011a).

### 13.6.2 Radio and optical observations of lightning during the GWS

Fig. 13.15 shows the number of detected SEDs as a function of time over nine months. Precisely speaking, it is the number of time-frequency events ("SED pixels") that is displayed in the figure for the following reason. Single SEDs normally extend over a few frequency channels, but high flash rates of 5-10 SEDs per second can lead to a temporal superposition of SEDs (Fischer et al., 2011a). As a consequence, single SEDs can no longer be resolved, because the most common receiver sweep rate is just 28 frequency channels per second (35.2 ms per channel). It can be seen in Fig. 13.15 that the number of SED pixels per Saturn rotation quickly (within 1-2 weeks) rose to values of $10^4$ to $10^5$, and remained at that level until mid-June 2011. This level is about an order of magnitude higher than the rate of regular (smaller) SED storms that typically emit $10^3$ to $10^4$ SEDs per Saturn rotation. Since the intensity of SEDs decreases with distance squared, the SED rate also depends on the distance of Cassini from Saturn. SEDs from regular SED storms and the 2010-2011 GWS have about the same spectral radio wave intensity; instead, only the SED rates are significantly different for these two classes of Saturn lightning storms (few SEDs per minute for regular storms and few SEDs per second for GWS). Flash rates of terrestrial thunderstorms are proportional to the horizontal area of the charge region (Larsen and Stansbury, 1974) suggesting that the charge region with convective activity was at least ten times larger for the GWS than for the regular, smaller Saturn lightning storms. The total number of SED pixels observed by the Cassini RPWS instrument during the great storm of 2010-2011 was almost 30 million.

Consistent with optical observations, the first SEDs of the GWS event were recorded on Dec. 5, 2010, and the last ones on Aug. 28, 2011. Fig. 13.15 shows a large decrease of the SED rate around June 20, 2011. This is most likely the consequence of the collision of the storm's head with the large anticyclonic vortex (section 13.3). The anticyclonic vortex in the tail had drifted westward with a rate of ~2°/day slower than the storm's head leading to the head-vortex collision as described in Section 13.3. Fig.13.16 shows the storm's head on the left side and the anticyclonic vortex in the tail on the right side, separated by ~120° in longitude at the end of February 2011. After the first large decrease in lightning activity on June 20, the storm regained some strength and showed intermittent strong SED activity around July 1, July 10, and July 20. For the rest of the time until the end of August, the SED activity was comparable to regular, smaller storms with almost no activity during mid-August. Some additional small SED activity not shown in Fig. 13.15 was present at the end of September/early October 2011 and at the end of December 2011. However, it is not clear where the source of this SED activity was located: either there was a small storm cell in the remaining visible clouds of the GWS at 35° N or in some other clouds at 50°N.

**[FIGURE 13.16]**

During the GWS the Cassini cameras made the first optical observation of Saturn lightning flashes on the day side (Dyudina et al., 2013). This was quite surprising, since it was even difficult to detect Saturn lightning flashes on the night side. Except at equinox, the light reflected from Saturn's rings causes the night side of Saturn to be brighter than Earth under a full moon, making it impossible to distinguish clouds illuminated from below by lightning from clouds illuminated from above by ring shine. Due to this, the first visible detection of lightning on Saturn was made on August 17, 2009, just a few days after Saturn northern vernal equinox (Dyudina et al., 2010), when the rings were edge-on to the Sun. The Cassini camera detected flash-illuminated cloud tops with diameters around 200 km on Saturn's night side at ~36° South latitude, accompanied by lightning-related radio emissions recorded by the Cassini RPWS instrument. The diameters of the bright spots suggested that the lightning is taking place 125-250 km below the cloud tops, which is above the base of the liquid $H_2O$-$NH_3$ (water-ammonia) cloud and most likely in the water ice cloud (Dyudina et al., 2010).

The lightning flashes during the GWS event were detected on the dayside in blue light only, leaving the lightning spectrum still unknown (Dyudina et al., 2013). The flashes could be detected on the dayside due to the high flash rate and by the technique of image subtraction. Since sunlit clouds do not change much within a few minutes, repeated images of the same area were used for background subtraction to reveal lightning. It turned out that the optical flashes occurred in the diagonal gaps between large anticyclones, which can be seen in Fig. 13.16. The head of the storm periodically spawned large anticyclones, which drifted off to the east with a longitudinal spacing of 10°-15°. The flashes marked by white crosses in Fig. 13.16 occur between them in cyclonic gaps, where the atmosphere looked clear down to the level of deep clouds (Dyudina et al., 2013). The flashes in Fig. 13.16 are distributed over a longitude range of ~70° and the western edge of the head at ~120° longitude is ~75° away from easternmost flashes at ~45°. The first SEDs in Fig. 13.16 are detected at 315° sub-spacecraft western longitude, which is consistent with an SED source at ~45° rising on the western dayside horizon. Due to Saturn's eastward rotation the easternmost SED sources are detected first before the storm's head rises on the horizon leading to an increase in SED activity. The longitudinal range of the optical flashes in Fig. 13.16 is thus consistent with the extended sub-spacecraft longitudinal range (~270°) over which SEDs were detected in late February 2011.

The optical energies of single Saturn lightning flashes were found to be typically ~ $10^9$-$10^{10}$ J, which is comparable to the total energy of a terrestrial flash (Dyudina et al., 2010; 2013). Since optical energies of flashes are just 0.1% of their total energy (Borucki and McKay, 1987), the total energy of an average Saturnian lightning flash should be in the range of $10^{12}$-$10^{13}$ J. This energy is comparable to the energy of a terrestrial *superbolt*. The vast amount of lightning energy goes into heating the channel and creating a shock wave. Fischer et al. (2011a) also estimated the total power of the GWS from the volume of the thundercloud and the latent heat release through condensation of water and arrived at a value of $10^{17}$ W, which is comparable to Saturn's total radiated power (~5 W/m²; corresponding to ~ 2.2 $10^{17}$ W over Saturn's surface area of 4.3 $10^{16}$ m² ). Dyudina et al. (2013) arrived at the same value via the following calculation. An average SED rate of a few flashes per second (peak rate >10 s$^{-1}$) for the GWS with a total energy of $10^{12}$-$10^{13}$ J per flash leads to a power of ~$10^{13}$ W released by lightning. The energetic efficiency of convective updrafts in producing lightning flashes is unknown for Saturn, but for Earth and Jupiter it is estimated to be around $10^{-4}$ (Borucki et al., 1982). Hence, a flash power of ~$10^{13}$ W for the SEDs of the GWS should be related to a total storm power of ~$10^{17}$ W, similar to the Fischer et al (2011a) result. This suggests that storms such as the 2010/2011 GWS are important players in the energy budget of Saturn's atmosphere. Furthermore, Fischer et al. (2014) suggested that the great storm of 2010-2011 might have even influenced the periodicity of Saturn kilometric radiation, an auroral radio emission which is modulated by Saturn's rotation. The GWS was most likely a source of intense gravity waves similar to terrestrial thunderstorms (Pierce and Coroniti, 1966) that may have caused a global change in Saturn's thermospheric winds via energy and momentum deposition. This supports the theory that Saturn's magnetospheric periodicities are driven by the planet's upper atmosphere.

## 13.7. Storm dynamics

As noted in section 13.2, the dynamics of a GWS can be divided in three steps that involve the (1) outbreak and (2) rapid growth to a large spot in interaction with the local wind, and (3) the zonal expansion to the planetary-scale disturbance. Here, we review our current understanding of the dynamics involved in each of these steps, as gleaned in particular from state-of-the-art computer modeling compared to the wealth of observations provided by Cassini and ground-based observers.

### 13.7. 1 The convective source

It is generally accepted that the outbreak of a GWS is due to moist convection in Saturn's cloudy weather layer (Sanchez-Lavega and Battaner, 1987; Hueso and Sanchez-Lavega, 2004). Thermochemical models predict the existence of three cloud types (NH$_3$-ice, NH$_4$SH-ice, H$_2$O-ice on top of an H$_2$O-NH$_3$ aqueous solution) that extend approximately between 1 and 10 bar depending on the abundances of N, O, S (Atreya and Wong, 2005; West et al., 2009). The convective hypothesis involves fast upward motions and is supported by the observed initial rapid evolution, fast expanding velocities with related horizontal divergence at cloud tops, high spot brightness (dense cloud containing fresh ices), lightning events, and spectroscopic ice detection at the cloud tops (section 13.4).

One-dimensional early parcel models of moist convection in Saturn showed that water located deep at 10-12 bar is the most effective fuel to heat the parcels through latent release and feed the updrafts, supported by an additional vertical driving given by ammonia at higher levels as the parcel ascends (Sanchez-Lavega and Battaner, 1987). Since water and ammonia have molecular weight greater than the ambient hydrogen atmosphere (by a factor ∼ 7-8), the ascent is favored when the parcel conditions are close to saturation with low entrainment of dry lighter environmental air into the parcel. Triggering moist convection under such conditions requires an initial heating of the parcel or a vertical initial velocity (see Hueso and Sanchez-Lavega, 2004). The 1D-models showed that for a highly enriched water atmosphere (5-10 times the solar oxygen abundance), the parcels are able to ascend the ∼ 260 km vertical distance from the water clouds to the tropopause located at the ∼ 100 mbar, reaching peak velocities up to ∼ 150 ms$^{-1}$. This can be seen to first order calculating the "Convective Available Potential Energy" CAPE given by (Sanchez-Lavega, 2011),

$$\text{CAPE} = \int_{z_1}^{z_2} g \frac{\Delta T}{T} dz = \frac{w_{max}^2}{2} \tag{1}$$

and then the maximum vertical velocity is given $w_{max} = \sqrt{2\text{CAPE}} \sim \sqrt{2g \frac{\Delta T}{T} h} \sim 150$ ms$^{-1}$ for $g = 10$ ms$^{-1}$, $T = 250$ K, $h = 260$ km, and assuming a temperature difference between the ascending parcel and the environment of just 1 K (Hueso and Sanchez-Lavega, 2004). When the updrafts reach the tropopause or the upper troposphere, overshooting could occur and mass flow conservation produces the horizontal expansion at the top and divergence in the velocity field. Vertical velocities in the diverging altitude can be simply computed by $w \le h \cdot \text{div} V$ where $\text{div}_h V = (1/A)(dA/dt) \approx 1\text{-}5\times10^{-6}$ s$^{-1}$ ($A$ is the area) as observed for the GWS 1933, 1990, 2010, yielding $w \le 2.5$ ms$^{-1}$.

Three-dimensional water and ammonia moist convective models were developed by Hueso and Sanchez-Lavega (2004). The model uses the anelastic approach and parameterized microphysics (Hueso and Sánchez-Lavega, 2001) in a mesoscale domain of few hundred kilometers (i.e., at the thermal plume scale) to study the trigger conditions and storm development. The most energetic thermal plumes are those resulting from water convection which reaches the 150 mbar altitude level developing peak vertical velocities of 150 ms$^{-1}$. When the atmosphere approaches instability, a very small temperature perturbation of $\Delta T = 0.15$ K is enough to initiate the onset. Because of the large path, the Coriolis force plays a major role in the thermal plume morphology.

Figure 13.17 shows a high resolution image of the 2010 GWS storm head and the measured wind field (García-Melendo et al., 2013). The convective source is the brightest area, spanning a meridional width ∼ 5,000 km and a zonal length of ∼ 2,500 km. Individual cumulus-like convective spots with ∼ 200 km in size and cumulus clusters can be observed. The interaction of the storm head with the westward jet where it resides generated a bow shock-like front with an arc-shaped morphology that approximately follows the zonal wind profile. However, the motions along this arc are intense with top velocities reaching 160 ms$^{-1}$ in an anticyclone open flow in gradient wind balance (García-Melendo et al., 2013). The radial pressure gradient was 2-5 times that of classical Earth mid-latitude baroclinic anticyclones and ∼ 2 times that of Jupiter's Great Red Spot, the largest anticyclone in the solar system. Numerical models that substitute the convective storm by a source injecting mass vertically are able to reproduce the observed morphology and dynamics under appropriate conditions in the weather layer. This requires that the static stability of the surrounding atmosphere measured in terms of the Brunt-Väisälä frequency (Sánchez-Lavega, 2011) should be low with $N \sim 1.7\times10^{-3}$ s$^{-1}$ to keep ambient conditions close to vertical instability. Importantly, to keep the disturbance morphology as observed, the source must inject mass continuously, which means that moist convection must be active for a long period. This is consistent with the continuously observed lightning activity (section 13.6). At the same time, the zonal winds must extend vertically across the weather layer (from ammonia to water clouds) but with a low wind shear $\partial u / \partial z \approx 5$ ms$^{-1}$/$H$ or lower with $H \sim 40\text{-}100$ km being the scale height.

**[FIGURE 13.17]**

### 13.7.2 The planetary scale disturbance

High resolution data obtained by the Cassini spacecraft on the Great Storm of 2010-2011 and the intensive ground-based coverage allowed, for the first time, to collect a unique wealth of data on the onset and evolution dynamics of a giant storm on Saturn (Sánchez-Lavega et al., 2011; Sánchez-Lavega et al., 2012; Sayanagi et al., 2013). The observed dynamical and morphological details provided a unique opportunity to perform numerical experiments to constrain Saturn's atmospheric structure in terms of fundamental variables such as vertical wind shear and vertical thermal structure. The 2010 event was simulated using the 3D *Explicit Planetary-Isentropic Coordinate* (EPIC) Atmospheric Model (Dowling et al., 1998), and shallow water (SW) models (García-Melendo and Sánchez-Lavega, 2016). By using these models, it was found that the most important morphological and dynamical features of the giant storm could be reproduced remarkably accurately, as we now discuss.

### 13.7.2.1. EPIC models of the Great Storm

The atmosphere of Saturn can be represented as a fluid under rotation whose motions are governed by classical geophysical fluid dynamics equations (Pedlosky, 1979; Sanchez-Lavega, 2011): the Navier-Stokes equation, the continuity equation, the state equation, and the first law of thermodynamics. Simplified versions of these equations valid under certain conditions met by the Great Storm allow numerical models to be calculated significantly faster without leaving out the essential physics. The EPIC model assumes that air parcel movements are adiabatic, or in other words the potential temperature obeys, $d\theta/dt = 0$ (an air parcel conserves its potential temperature). As a consequence, the Ertel potential vorticity ($PV$)

$$PV = \frac{(\vec{\omega} + 2\vec{\Omega})}{\rho} \bullet \nabla \theta \qquad (2)$$

is also a conserved quantity where $\vec{\omega} = \nabla \times \vec{u}$ is the parcel vorticity vector ($\vec{u}$ is the parcel 3D velocity on the planet), $\vec{\Omega}$ is the rotation vector whose modulus is the planet's rotation angular velocity and whose direction follows the planet's rotation axis in the south-north direction, and $\rho$ is fluid's density. Since potential vorticity is conserved by air parcels, it can be used as a passive tracer to represent atmospheric movements. In addition, EPIC uses isentropic coordinates where the potential temperature substitutes for the natural vertical coordinate *z*. Under this choice, the vertical domain can be divided in layers of constant $\theta$ where movements are purely horizontal ($\dot{\theta} = d\theta/dt \equiv 0$). The adiabatic approximation has proven to be very useful to model meteorological features in the giant planets (e. g. LeBeau and Dowling, 1998; Garcia-Melendo et al, 2007; Legarreta and Sánchez-Lavega, 2008; Sayanagi et al., 2010).

The model needs a set of free parameters to specify the atmospheric structure. The most important are the zonal wind profile *U(y)*, the vertical wind shear *U(z)*, and the vertical static stability *N(z)*. The simulated vertical domain extended from the stratosphere at 10 mbar, down to the water condensation level at ~ 10 bar where the storm originated (Sanchez-Lavega and Battaner, 1987; Hueso and Sanchez-Lavega, 2004).

There is precise information on $U(y)$ at the visible cloud deck (~ 500 mbar) in the latitude band where the 2010 GWS appeared before the storm's onset. Since orbital insertion in 2004, Cassini has monitored the global zonal circulation at different wavelengths, in particular establishing its stability during the time preceding the storm's outbreak between 2004 and 2011 (Porco et al., 2005; Vasavada et al., 2005; García-Melendo et al., 2011). The vertical thermal profile is also well known above the visible cloud deck down to ~ 1 bar. It was retrieved by the radio occultation experiments performed during the Voyager mission flybys (Lindal et al., 1985) down to ~ 1 bar, and later corrected for the updated helium molar fraction of 0.13 (Conrath and Gautier, 2000) which was not well determined at the time of the Voyager experiments. Vertical wind shear is mostly unknown under the visible cloud deck. Assuming that the 2010 GWS source is at the 10-bar water-condensation level and its observed velocity is rooted at this depth, then the winds must strengthen with depth (Sánchez-Lavega et al., 2011; Sánchez-Lavega et al., 2012), or at least maintain its intensity. Therefore, Sánchez-Lavega et al. (2011) 3D simulations tested constant and slightly increasing winds with depth.

The vertical thermal profile is also unknown at deep levels under the visible cloud deck. The corrected thermal profile is somewhat unstable for p > 400 mbar (Brünt-Väisälä frequency $N^2 < 0$), but due to the presence of convective activity and vortices, stable and unstable conditions can coexist (Sánchez-Lavega et al., 2003, 2004; Porco et al., 2005). EPIC does not support convective motions, therefore the simulations assume that the atmosphere is statically stable. This is a realistic assumption for atmospheric levels above over the tropopause, and a reasonable approximation for deeper levels, if low static stability values are used, taking in mind that static stability may oscillate around a neutral value and likely approaches it as we get closer to the 10 bar level where $N^2 \approx 0$. Vertical thermal profiles were tested in the range $0.03 \times 10^{-4} \, s^{-2} \leq N^2 \leq 0.7 \times 10^{-4} \, s^{-2}$ for the deepest levels (Sánchez-Lavega et al. 2011).

### 13.7.2.2 The Shallow Water model

The relevant fluid equations can be simplified further by assuming that atmospheric movements take place in a relatively thin sheet of fluid compared to the much larger horizontal scale of movements, and that the atmospheric density is spatially constant within each pressure level. The first assumption is a very reasonable one if we consider that there is a ~ 260 km excursion from the water condensation level at ~ 10 bar to the tropopause (Hueso and Sanchez-Lavega, 2004), much smaller than the ~$10^4$ - $10^5$ km horizontal scale covered by the giant storm. Assuming lack of vertical stratification may seem a poor approximation, but the model is still able to capture much of the essential physics, and in many situations the atmosphere behaves as an incompressible fluid when movements are horizontal. There are several consequences under these assumptions: thermodynamics and dynamics become uncoupled, the system becomes hydrostatic, and movements become two dimensional. Carrying out these approximations, we obtain the Shallow Water (SW) equations which form a much more simplified version of the Navier-Stokes equations (Vallis, 2006):

$$\frac{d\vec{u}}{dt} + f\vec{k} \times \vec{u} = -g\nabla h \qquad (3)$$

$$\frac{\partial h}{\partial t} = -\nabla \bullet (h\vec{u}) + S(\vec{r},t). \qquad (4)$$

Equations (3) and (4) are the respective momentum and continuity equations, $\vec{u} = (u,v)$ is the horizontal velocity field, with $u$ and $v$ the respective zonal and meridional velocities, d/dt is the temporal derivative, $f = 2\Omega\sin\varphi$ is the Coriolis parameter (where $\varphi$ is the latitude), $h$ is the model's total layer thickness, $\vec{r}$ is the vector position, g is the local gravity, $S$ is the source of mass which models the storm, and $\vec{k}$ is the unit vector in the vertical direction. SW models have been previously used in the study of dynamics on giant planet in both our and other solar systems (Vasavada and Showman, 2005). In all cases the upper lower density layer represents the weather layer under study, and a bottom denser layer mimicks the deep interior of the planet. We performed 1-layer and 2-layer SW models to simulate the response of Saturn's atmosphere to the massive storm onset. Topography, after applying geostrophic equilibrium, may introduce the zonal wind profile, but a flat bottom may equally well work if the effect of $U(y)$ is numerically included during the computation of the prognostic variables in the form $U(y)+u$. In this way, $U(y)$ is maintained fixed, but it will let $u$ evolve freely under the influence of zonal winds. Rayleigh damping was also introduced to prevent long-term turbulence.

### 13.7.2.3 The storm source

A fundamental observed aspect of the Great Storm of 2010-2011 is its strong, continuous activity over at least a 7- month period.  Dynamically, the convective activity within the storm head appeared to cease when it met the long-lived vortex (Sánchez-Lavega at al., 2012). However, SED lightning detections indicate that there was additional convective activity within the storm head or nearby for about 2 months longer (Sayanagi et al., 2013), perhaps indicating strong dynamics at depth under the cloudtops. It is therefore fundamental for modellers to simulate long-lived convective storm systems in numerical experiments, in order to adequately reproduce the response of Saturn's atmosphere.

The EPIC modeling allows an injection of mass and energy in the form of heat in all or part of the vertical domain. Since SW models are totally decoupled from the thermodynamic processes, the storm's action on the atmosphere is modelled as a combined effect of mass and energy perturbations in the form of a perturbed surface elevation (García-Melendo et al., 2013). For both models a continuously active Gaussian pulse is used for the perturbation injection function *S*. In the SW model, a bump with a Gaussian shape is injected in the surface of the upper layer only (representing Saturn's weather layer) at the storm location, where the storm energy and mass are deposited interacting with the background zonal winds. In the EPIC model we introduce a spatial Gaussian heat distribution around the injection point. Following observations, storm's horizontal size is limited to a circular patch about 1°-2° in radius, with a spatial Gaussian decay. We moved the simulated perturbation at the zonal velocity of ~28 m s$^{-1}$, the same one of the real storm. It must be noted that both the SW and EPIC models are approximations to the real atmosphere; the perturbation parameters are introduced *ad hoc*, tuned until an adequate response of the model atmosphere is obtained, which then yields clues on the storm's dynamics. As a consequence, it is not possible to derive absolute realistic values for the amount of energy and mass injected by the storm. In order to infer realistic values, modellers need to know something about the source of the storm and implement it in a more realistic model which includes convection, cloud formation, latent heat release due to phase changes of condensable species, etc.

### 13.7.2.4. Simulation results

Both the EPIC and SW models yielded relevant information about the structure of Saturn's atmosphere and the dynamics of the 2010 perturbation (Sanchez-Lavega et al, 2011; García-Melendo and Sanchez-Lavega, 2015). One important result was the inability to reproduce the observed morphology of the storm unless a specific combination of parameters initiated the calculations. For instance, they found that inadequate latitude injection within a narrow range of +/-2° latitude always led to a morphology pattern totally different from that observed. In the case of the SW model, good results were obtained only when the magnitude of the Rossby deformation radius L$_R$ was within certain limits (300 km ≤ $L_R$ ≤ 1000 km). For large $L_R$, the model turned out to be unstable with the production of large vortices (Showman, 2007). In the end, Sanchez-Lavega et al (2011) and Garcia-Melendo and Sanchez-Lavega (2016) found they could reproduce the dynamics and morphology quite well, giving them confidence in the simulation results. In the 3D EPIC models, they obtained a PV field that was best able to simulate the storm with almost constant or slightly increasing winds below the cloud deck and low values of static stability, $N^2 = 0.03 \times 10^{-4}$ s$^{-2}$, specially during its onset (Figure 13.18). Such an atmospheric structure is also consistent with previous simulation results (García-Melendo et al., 2007; del Rio-Gaztelurrutia et al., 2007).

**[FIGURE 13.18]**

The SW and 3D numerical experiments of Sanchez-Lavega et al (2011) and Garcia-Melendo and Sanchez-Lavega (2016) revealed the dynamic mechanism in the storm's head that drove the development of most of the remainder of the storm, such as the evolution of the tail and the north and south branches. According to the simulations, the head of the storm was adequately reproduced only when strong relative anticyclonic vorticity, interacting with the prevailing zonal winds, was injected into the atmosphere after the addition of mass or heat (see Dyudina et al., 2013). In turn, this indicates that convection lifted enormous amounts of both heat and mass. This is fully consistent with the observations (Sanchez-Lavega et al., 2011). Neither EPIC nor SW models can simulate convection, but geostrophic adjustment can be reproduced in both models. The effect is dramatic: as well produced in the 3D and SW models (Figure 13.2), when the perturbation moves at a speed faster that the local winds (-29 m s$^{-1}$ for the 2010 GWS), in a background with strong meridional wind shear of the zonal wind profile, a bow front at the storm's head appears with intense anticyclonic circulation, $\sim 160$ ms$^{-1}$ along its periphery (García-Melendo et al., 2013) constituting the tail up to the observed long-lived anticyclonic vortex (Dyudina et al., 2013). According to these model analyses, anticyclonic circulation is also responsible for shedding the long-lived anticyclone, during the onset of the storm, which drifted to the end of the storm's tail (Sánchez-Lavega et al. 2012; Sayanagi et al., 2013). To survive as a compact anticyclone, it translated to the north to rest entirely within the anticyclonic flank of the $\sim$ 50ºN jet, at $\sim$41ºN.

**[FIGURE 13.19]**

Morphologically important details are also reproduced by numerical experiments (Figures 13.18 and 13.19). The frontal-arc shaped structure forms at the beginning of the storm, reproducing very well the observed initial stages of the storm (compare Figs. 13.5 and 13.17 with Figs. 13.18 and 13.19). One interesting aspect captured by the numerical models is the quick expansion of part of the storm material towards the south, forming a turbulent south branch with chains of small vortices. Rapid expansion towards the south is due to the background relative vorticity where the storm was embedded. The 2010 event erupted near the peak of the westward jet centred at planetographic latitude $\sim$ 40ºN (Sánchez-Lavega et al., 2012), just in between the cyclonic north flank of the equatorial jet, and the anticyclonic north part of the $\sim$ 48ºN eastward jet. It is well known that vortices survive in regions with the same relative vorticity, and they are rapidly sheared apart when the ambient vorticity is the opposite (Ingersoll et al., 1984). Simulations show that the particular point where the 2010 GWS erupted is responsible for its quick dispersion on its south side. The highly anticyclonic material injected by the storm was in part quickly sheared and dispersed southwards of $\sim$ 38ºN, forming wavy patterns which turned into a turbulent region. Simulations show that the chain of small vortices are little anticyclones formed at the 34ºN to 30ºN planetographic latitude interval (Sánchez-Lavega et al., 2012), a region where the undisturbed zonal wind profile presented two small inflexions. These vortices disappeared later, perhaps due to the alteration of the zonal profile in this region at the visible cloud top level ($\sim$ 500 mbar), which smoothed out the zonal wind field (Fig. 13.6, Sayanagi et al., 2013). On the other hand, the material injected by the storm reaching the anticyclonic side of the profile was accelerated to reinforce the anticyclonic cells and accelerate the material on the edge of the northern part of the storm head.

Sayanagi et al. (2013) observed that the local zonal wind speed around the 2010 storm eruption latitude had changed by as much as ~ 40 m s$^{-1}$ (Figure 13.6). The authors proposed two possible mechanisms for such a change: relative vorticity injection due to convective cloud divergence under rotational conditions, and a change in the latitudinal gradient of temperature due to latent heat release, which would change vertical wind shear and therefore the observed cloud top velocities due to the thermal wind effect. Relative vorticity injection is well reproduced in SW models. Figure 13.20 shows how the zonal wind profile is affected by the injection of vorticity according to simulations. The change is similar to that measured in real images (compare Figs. 13.6 and 13.20), confirming that relative vorticity injection may, in part, alter zonal wind speed.

**[FIGURE 13.20]**

### 13.7.3 The recurrence interval

Several authors have addressed the intermittency between convective events on Jupiter and Saturn. Guillot (1995) was the first to point out that hydrogen gas saturated with water or ammonia vapor will have a density minimum as a function of temperature, provided the mixing ratios of the vapors are high enough. For instance, a parcel saturated with water vapor and a mixing ratio greater than 1% will first become less dense as it cools. This is because the heavier water vapor precipitates out, which overwhelms the normal thermal contraction due to cooling. A mixing ratio of 1% is about 10 times the abundance in a solar composition atmosphere. Such 10-fold enrichment is suspected for Saturn but not for Jupiter, based largely on methane and phosphine, two gases whose abundances have been inferred spectroscopically.

Li and Ingersoll (2015) applied this mass-loading effect to Saturn. After a giant storm, which is treated as a giant convective event, the troposphere is warm and saturated. As it cools, it becomes less dense, and a stable interface develops at cloud base. The cooling takes decades, due to the long radiative time constant of Saturn's troposphere—it has high mass per unit area and it radiates to space at a low temperature. When the troposphere has unloaded most of its water vapor, further cooling makes the troposphere denser, which erodes the stable interface until another giant convective event occurs. Li and Ingersoll model the convection and the slow cooling using a 1D radiative-convective equilibrium model, and they model the geostrophic adjustment following a convective event using a 2D axisymmetric dynamical model. The latter shows how the atmosphere can be sub-saturated with respect to ammonia after a giant storm (Janssen et al. 2013, Laraia et al. 2013).

Sugiyama et al. (2011, 2014) have a 2D cloud-resolving model that explicitly represents the convective motion and the microphysics of the three cloud components $H_2O$, $NH_3$, and $NH_4SH$. In their model, stable layers associated with condensation develop and act as dynamical barriers to convection. As cooling proceeds, the moist convective layer becomes potentially unstable and is triggered by the updrafts from below that are forced by neighboring downdrafts, which themselves are forced by the cooling effect of falling raindrops as they evaporate. Sugiyama et al. find that the period of intermittency is roughly equal to the time obtained by dividing the mean temperature increase, which is caused by active cumulonimbus development, by the body cooling rate. In essence, this means that the radiative time constant is setting the long time interval between giant storms. Although the mechanisms are different, the models of Sugiyama et al. and Li and Ingersoll agree on this important point.

Alternatively, the long recurrence interval between GWSs, of the order of the Saturn year, could be controlled by the seasonal insolation heating and cooling cycle that suffers the upper cloud and hazes (see Figure 13.1). Heating rates in Saturn's upper atmosphere have been calculated by Perez-Hoyos and Sanchez-Lavega (2006) show that the solar flux is deposited above ~ 200 mbar. It remains to be seen if the solar insolation cycle affects the deep cloudy layers to trigger convection on them.

**[FIGURE 13.21]**

## 13.8. Conclusions

As the most recent and best-studied example of a GWS, The Great Storm of 2010-2011 represents a unique and major dynamical phenomenon in planetary atmospheres Figure 13.21 summarizes schematically our ideas of the structure and major dynamical drivers of the 2010-11 GWS. As a class of meteorological events that can be used to elucidate comparative planetology, the GWS can be considered as a testing bench of the state and structure of the atmospheres of the giant planets or, more generally, of hydrogen-dominated atmospheres. Their importance transcends the mere study of huge, quasi-periodic dynamical disturbances on Saturn. They involve a large spectrum of dynamical phenomena: moist convection, a variety of waves, vortex generation, turbulence etc., at all the scales of motion, from mesoscale to planetary scale. Their study yields information about the thermal structure, composition (for example oxygen and water), compound generation and their transport, and microphysics (particle formation) from the troposphere to the stratosphere. They also involve the study of electrical phenomena (lightning) in the clouds of the giant planets. They provide insights into the nature and vertical structure of the zonal jets in giant planets and how they respond to the outbreak of huge disturbances. Finally, such energetic storms – in the case of the 2010-2011 Great Storm, emitting a significant fraction of a planet's total radiated power for many months - could also play an important role in the planet's global energy balance.

The 2010-11 event has been a breakthrough in the understanding of the GWS but there are still a large number of major unknowns about the phenomenon. How, exactly, periodic are they, and is this potential periodicity a key to explaining their power? Do they occur in the southern hemisphere? If not, then what is the reason for their concentration in the northern hemisphere taking into account the hemispheric symmetry in the insolation cycle? Can they outbreak at any latitude? How are they triggered? And how is their immense power maintained over many months? As well, a plethora of physical-chemical phenomena produced by GWS events in the troposphere and stratosphere remain to be explained, including the formation of powerful, high-temperature "*beacons*" in the high stratosphere. Particularly important are the Equatorial events for which we have relatively sparse information. A major difficulty in answering these questions through the acquisition of new data is the rarity of the phenomenon, historically occurring about every three decades. Therefore it is important to exploit to the maximum all the information that Cassini has obtained and is still obtaining on Saturn, and together with it, to develop more advanced models than those thus far employed, to fully explain the phenomenon and its global implications on the structure and dynamics of Saturn's atmosphere.

## Acknowledgments

A. Sanchez-Lavega, E. Garcia-Melendo and S. Perez-Hoyos are supported by the Spanish project AYA2012-36666 with FEDER support, Grupos Gobierno Vasco IT-765-13 and by Universidad del País Vasco UPV/EHU through program UFI11/55. L. Sromovsky is supported for this work by NASA through its Outer Planets Research Program under grant NNX11AM58G. L. Fletcher was supported by a Royal Society Research Fellowship at the University of Leicester. G. Fischer was supported by the Austrian Science Fund FWF through project P24325-N16. B.Hesman was supported by the NASA Cassini/CIRS project, by the NASA Planetary Astronomy Program (PAST) under grant number NNX11AJ47G, and by the NASA Cassini Data Analysis and Participating Scientists Program (CDAPS) under grant number NNX12AC24G. K. M. Sayanagi was supported in part by grants from

# References

Acarreta, J.R., A, and Sánchez-Lavega, A. (1999). Vertical cloud structure in Saturn's 1990 Equatorial storm. *Icarus*, **137**, 24–33. Doi: 10.1006/icar.1998.6034.

Achterberg, R. K., J., Gierasch P., Conrath, B. J., et al. (2014). Changes to Saturn's zonal mean tropospheric thermal structure after the 2010-2011 northern hemisphere storm. *Astrophysical Journal*, **786**, 92, pp8. Doi: 10.1088/0004-637X/786/2/92.

Atreya, S. K., and Wong, A.-S. (2005). Coupled clouds and chemistry of the giant planets – a case for multiprobes. *Space Science Reviews*, **116**, 121-136. Doi: 10.1007/s11214-005-1951-5.

Baines, K. H., Delitsky, M. L., Momary, et al. (2009). Storm clouds on Saturn: Lightning-induced chemistry and associated materials consistent with Cassini/VIMS spectra. *Planetary and Space Sci.* **57**, 1650-1658. doi:10:1016/j.pss.2009.06.025

Baines, K.H., Fletcher, L. N, Momary, T. W., et al. (2015). The Evolution of Saturn's String of Pearls over Five Years as Revealed by Cassini/VIMS, (under review at Icarus).

Baines, K. H., Momary, T. W., Fletcher, L. N., et al. (2010). Saturn's "String of Pearls" After Five Years: Still There, Moving Backwards Faster in the Voyager System. *Bulletin of the American Astronomical Society*, **42**, 1039.

Barnet, C.D., Westphal, J.A., Beebe, R.F., and Huber, L.F. (1992). Hubble space telescope observations of the 1990 Equatorial disturbance on Saturn: Zonal winds and central meridian albedos. *Icarus*, **100**, 499–511. Doi: 10.1016/0019-1035(92)90113-L.

Beebe R. F., Barnet, C., Sada, P. V., and Murrell, A. S. (1992). The onset and growth of the 1990 equatorial disturbance on Saturn, *Icarus*, **95**, 163 - 172. Doi: 10.1016/0019-1035(92)90035-6.

Bézard, B., Moses, J. I., Lacy, J., et al. (2001). Detection of Ethylene ($C_2H_4$) on Jupiter and Saturn in Non-Auroral Regions. *Bulletin of the American Astronomical Society*, **33**, 1079.

Bjoraker, G., Hesman, B. E., Achterberg, R. K., and Romani, P. N. (2012). The Evolution of Hydrocarbons in Saturn's Northern Storm Region. *Bulletin of the American Astronomical Society*, 44, #403.05.

Brown, R. H., Baines, K. H., Bellucci, G., et al. (2004). The Cassini Visual and Infrared Mapping Spectrometer (VIMS) Investigation. *Space Science Reviews* **115**, 111–168. Doi: 10.1007/s11214-004-1453-x.

Borucki, W. J., Bar-Nun, A., Scarf, F. L., Look, A. F., and Hunt, G. E. (1982), Lightning activity on Jupiter, *Icarus*, **52**, 492-502. Doi: 10.1016/0019-1035(82)90009-4.

Borucki, W. J., and C. P. McKay (1987), Optical efficiencies of lightning in planetary atmospheres, *Nature*, **328**, 509-510. Doi: 10.1038/328509a0.

Cavalié, T., Dobrijévic, M., Fletcher, L. N., et al. (2015). The photochemical response to the variation of temperature in Saturn's 2011-2012 stratospheric vortex. *Astronomy & Sstrophysics,* 580, A55

Charney, J. G., and Drazin, P. G. (1961). Propagation of planetary-scale disturbances from the lower into the upper atmosphere. *Journal of Geophysical Research*, **66**, 83-109. Doi: 10.1029/JZ066i001p00083.

Choi, D. S., Showman, A. P., and Brown, R. H. (2009). Cloud features and zonal wind measurements of Saturn's atmosphere as observed by Cassini/VIMS. *Journal of Geophysical Research*, **114**, E04007. Doi: 10.1029/2008JE003254.

Clark, R. N., and 13 colleagues (2012). The surface composition of Iapetus: Mapping results from Cassini VIMS. Icarus ,218, 831-860.

Conrath, B.J., and Gautier, D. (2000). Saturn helium abundance: A reanalysis of Voyager measurements, *Icarus*, **144**, 124–134, Doi: 10.1006/icar.1999.6265.

del Río-Gaztelurrutia, T., J., Legarreta, R., Hueso, S., Pérez-Hoyos, and A. Sánchez-lavega, (2010). A long-lived cyclone in Saturn's atmosphere: Observations and models, *Icarus*, **209**, 665–681, doi: 10.1016/j.icarus.2010.04.002

Dowling, T.E., Fischer, A.S., Gierasch, P.J., Harrington, J., Lebeau, R.P., and Santori, C.M. (1998). The Explicit Planetary Isentropic-Coordinate (EPIC) atmospheric model. *Icarus*, **132**, 221–238, doi: 10.1006/icar.1998.5917.

Dyudina, U. A., Ingersoll, A. P., Ewald, S. P., et al. (2007). Lightning storms on Saturn observed by Cassini ISS and RPWS during 2004-2006. *Icarus*, **190**, 545-555, doi:10.1016/j.icarus.2007.03.035.

Dyudina, U.A., Ingersoll, A. P., Ewald, S. P., et al. (2010). Detection of visible lightning on Saturn. *Geophysical Research Letters*, **37**, L09205. Doi: 10.1029/2010GL043188.

Dyudina, U.A Ingersoll, A. P., Ewald, S. P., Porco, C. C., Fischer, G. and Yair, Y. (2013). Saturn's visible lightning, its radio emissions, and the structure of the 2009-2011 lightning storms. *Icarus*, **226**, 1020-1037, doi:10.106/j.Icarus.2013.07.013.

Elachi C., Allison, M. D., Borgarelli, L, et al. (2004). Radar: The Cassini Titan Radar Mapper. *Space Science Reviews*, **115**, 71-110. Doi: 10.1007/s11214-004-1438-9.

Encrenaz, T., Combes, M., Zeau, Y., Vapillon, L., and Berezne, J. (1975). A Tentative Identification of $C_2H_4$ in the Spectrum of Saturn. *Astronomy & Astrophysics*, **42**, 355-356.

Fischer, G., Desch, M. D., Zarka, P., et al. (2006). Saturn lightning recorded by Cassini/RPWS in 2004. *Icarus*, **183**, 135-152, Doi:10.106/j.icarus.2006.02.010.

Fischer, G., Kurth, W. S., Dyudina, U. A., et al., (2007a), Analysis of a giant lightning storm. *Icarus*, **190**, 528-544, Doi:10.1016/j.icarus.2007.04.002.

Fischer, G., Gurnett, D. A., Lecacheux, A., Macher, W., and Kurth, W. S. (2007b). Polarization measurements of Saturn Electrostatic Discharges with Cassini/RPWS below a frequency of 2 MHz, *Journal of Geophysical Research*, **112**, A12308, Doi:10.1029/2007JA012592.

Fischer, G., Gurnett, D. A., Kurth, W. S., et al. (2008). Atmospheric Electricity at Saturn, *Space Science Review*, **137**, 271-285, Doi:10.1007/s11214-008-9370-z

Fischer, G., Kurth, W. S., Gurnett, D. A., et al. (2011a). A giant thunderstorm on Saturn. *Nature*, **475**, 75-77. Doi: 10.1038/nature10205.

Fischer, G., Gurnett, D.A., Zarka, P., Moore, L., and Dyudina, U. A., (2011b). Peak electron densities in Saturn's ionosphere derived from the low-frequency cutoff of Saturn lightning. *Journal of Geophysical Research*, **116**, A04315, doi:10.1029/2010JA016187.

Fischer, G., Dyudina, U. A., Kurth, W. S., et al. (2011c). Overview of Saturn lightning observations, in: Planetary Radio Emissions VII, edited by H.O. Rucker, W.S. Kurth, P. Louarn, and G. Fischer, Austrian Academy of Sciences Press, Vienna, 135-144.

Fischer, G., Ye, S.-Y., Groene, J.B., et al., (2014). A possible influence of the Great White Spot on Saturn kilometric radiation periodicity. *Annales Geophysicae*, **32**, 1463-1476. Doi:10.5194/angeo-32-1463-2014. Doi: 10.5194/angeo-32-1463-2014.

Flasar F. M. Kunde, V. G.; Abbas, M. M., et al. (2004). Exploring The Saturn System In The Thermal Infrared: The Composite Infrared Spectrometer. *Space Science Reviews*, **115**, 169-297. Doi: 10.1007/s11214-004-1454-9.

Fletcher, L. N., Hesman, B. E., Irwin, P. G., at al. (2011). Thermal Structure and Dynamics of Saturn's Northern Springtime Disturbance. *Science*, **332**, 1413-1417. Doi: 10.1126/science.1204774.

Fletcher, L. N., Hesman, B. E., Achterberg, R. K., et al. (2012). The origin and evolution of Saturn's 2011-2012 stratospheric vortex. *Icarus*, **221**, 560-586. Doi: 10.1016/j.icarus.2012.08.024.

Fletcher, L. N., Greathouse, T. K., Orton, G. S., et al. (2014). The origin of nitrogen on Jupiter and Saturn from the 15N/14N ratio. *Icarus*, **238**, 170-190. Doi: 10.1016/j.icarus.2014.05.007.

García-Melendo, E., Sánchez-Lavega, A., and Hueso, R. (2007). Numerical models of Saturn's long-lived anticyclones. *Icarus*, **191**, 665-677. Doi: 10.1016/j.icarus.2007.05.02

García-Melendo, E., Pérez-Hoyos, S., Sánchez-Lavega, A., and Hueso, R. (2011). Saturn's zonal wind profile in 2004-2009 from Cassini ISS images and its long-term variability. *Icarus*, **215**, 62-74. Doi: 10.1016/j.icarus.2011.07.005.

García-Melendo, E., Hueso, R., Sánchez-Lavega, A., et al. (2013). Atmospheric dynamics of Saturn's 2010 giant storm, *Nature Geoscience*, **6**, 525-529. Doi: 10.1038/NGEO1860.

García-Melendo, E., and Sánchez-Lavega, A. (2016). Shallow Water simulations of Saturn's Great White Spots at different latitudes. *Icarus,* accepted. Doi: 10.1016/j.icarus.2016.10.006.

Guerlet, S., Spiga, A., Sylvestre, M., et al. (2014). Global climate modeling of Saturn's atmosphere. Part I: Evaluation of the radiative transfer model. *Icarus*, **238**, 110-124. Doi: 10.1016/j.icarus.2014.05.010.

Guillot, T. (1995). Condensation of methane, ammonia, and water and the inhibition of convection in giant planets. *Science,* **269,** 1697- 1699. Doi:10.1126/science.7569896

Gurnett, D.A., W.S. Kurth, D.L. Kirchner, et al. (2004). The Cassini radio and plasma wave science investigation. Space Science Reviews, 114, 395-463. Doi: 10.1007/s11214-004-1434-0.

Harvey, V. L., M. H. Hitchman (1996). A Climatology of the Aleutian High. *Journal of Atmospheric Sciences*, 53, 2088 – 2102.

Harvey, V. L., Pierce, R. B., Hitchman, M. H., Randall, C. E., and Fairlie, T. D. (2004). On the distribution of ozone in stratospheric anticyclones. *Journal of Geophysical Research (Atmospheres)*, **109**, 24308. Doi: 10.1029/2004JD004992.

Hesman, B.E., G.L. Bjoraker, P.V. Sada, et al. (2012). Elusive Ethylene Detected in Saturn's Northern Storm Region. *The Astrophysical Journal*, **760**, 24-30. Doi: 10.1088/0004-637X/760/1/24.

Hesman, B.E., G.L. Bjoraker, R.K. Achterberg et al. (2013). The Evolution of Hydrocarbon Compounds in Saturn's Stratosphere During the 2010 Northern Storm. *AGU Fall Meeting Abstracts*, *Dec. 2013*, **C2205**.

Hesman, B. E., Bjoraker, G. L., Sada, P. V., et al. (2012). Elusive Ethylene Detected in Saturn's Northern Storm Region. *Astrophys. J.,* **760**, 24 (7pp). Doi:10.1088/0004-637X/760/1/24

Hueso R., and Sánchez-Lavega, A. (2001). A three-dimensional model of moist convection for the giant planets: The Jupiter case. *Icarus*, **151**, 257-274. Doi: 10.1006/icar.2000.6606.

Hueso, R., Sánchez-Lavega, A. (2004). A three-dimensional model of moist convection for the giant planets II: Saturn's water and ammonia moist convective storms. *Icarus*, **172**, 255-271. Doi: 10.1016/j.icarus.2004.06.010.

Hueso, R., Legarreta, J., Pérez-Hoyos, S., Rojas, J F., Sánchez-Lavega, A., and Morgado, A. (2010). The international outer planets watch atmospheres node database of giant-planet images. *Planetary and Space Science*, **58**, 1152-1159. Doi:10.1016/j.pss.2010.04.006.

Hurley, J., Irwin, P. G. J., Fletcher, et al. (2012). Observations of upper tropospheric acetylene on Saturn: No apparent correlation with 2000 km-sized thunderstorms. *Planetary and Space Science*, **65**, 21-37. Doi: 10.1016/j.pss.2011.12.026.

Ingersoll, A. P., Beebe, R. F., Conrath, B. J., and Hunt, G. E. (1984). Structure and dynamics of Saturn's atmosphere. Saturn. Gehrels, T., and M. S. Matthews, *University Of Arizona Press*, 195–238.

Janssen, M. A., Ingersoll, A. P., Allison, M. D., et al. (2013). Saturn's thermal emission at 2.2-cm wavelength as imaged by the Cassini RADAR radiometer. *Icarus*, **226**, 522-535. Doi: 10.1016/j.icarus.2013.06.008.

Kaiser, M. L., M. D. Desch, and Connerney, J. E. P. (1984). Saturn's ionosphere: Inferred electron densities. *Journal of Geophysical Research*, **89**, A4, 2371-2376. Doi: 10.1029/JA089iA04p02371.

Karkoschka, E., and Tomasko, M. G. (1993). Saturn's upper atmospheric hazes observed by the Hubble space telescope. *Icarus*, **106**, 428–441. Doi: 10.1006/icar.1993.1183.

Karkoschka, E., and Tomasko, M. G. (2005). Saturn's vertical and latitudinal cloud structure 1991–2004 from HST imaging in 30 filters. *Icarus* **179**, 195–221. Doi: 10.1016/j.icarus.2005.05.016.

Konovalenko, A. A., Kalinichenko, N. N., Rucker, H. O. et al. (2013). Earliest recorded ground-based decameter wavelength observations of Saturn's lightning during the giant E-storm detected by Cassini spacecraft in early 2006, *Icarus*, **224**, 14-23, doi:10.1016/j.icarus.2012.07.024.

Laraia, A. L., Ingersoll, A. P., Janssen, M. A., Gulkis, S., Oyafuso, F., and Allison, M. (2013). Analysis of Saturn's thermal emission at 2.2-cm wavelength: Spatial distribution of ammonia vapor. *Icarus*, **226**, 641-654. Doi: 10.1016/j.icarus.2013.06.017.

Larsen, H.R., and Stansbury, E. J. (1974). Association of lightning flashes with precipitation cores extending to height 7 km. *Journal of Atmospheric and Terrestrial Physics*, **36**, 1547-1533.

Lebeau, R.P., and Dowling, T. E. (1998). EPIC simulations of time-dependent, three-dimensional vortices with application to Neptune's great dark spot. *Icarus*, **132**, 239–265. Doi: 10.1006/icar.1998.5918.

Legarreta, J., and Sánchez-Lavega, A. (2008). Vertical structure of Jupiter's troposphere from nonlinear simulations of long-lived vórtices, *Icarus*, **196**, 184–201. Doi: 10.1016/j.icarus.2008.02.018.

Li, C. and Ingersoll, A. P. (2015). Moist convection in hydrogen atmospheres and the frequency of Saturn's giant storms. *Nat. Geoscience,* **8**, 398-403. doi:10.1038/ngeo2405

Li L., Jiang X., Trammell H. J., Pan Y., Hernandez J., Conrath B. J., Gierasch P. J., Achterberg R. K., Nixon C. A., Flasar F. M., Perez-Hoyos S., West R. A., Baines K. H., and Knowles B. (2015). Saturn's giant storm and global radiant energy. *Geophys. Res. Lett.,* **42**, 2144-2148. doi: 10.1002/2015GL063763.

Lindal, G.F., Sweetnam, D. N., and Eshleman, V. R. (1985). The atmosphere of Saturn: An analysis of the Voyager radio occultation measurements. A*stronomical Journal*, **90**, 1136–114. doi: 10.1086/113820.

Manney, G. L., Froidevaux, L., Waters, J. W., et al. (1995). Formation of low-ozone pockets in the middle stratospheric anticyclone during winter. *Journal of Geophysical Research-Atmospheres*, **100**, 13939-13950. Doi: 10.1029/95JD00372.

Miller, E. A., Klein, G., Juergens, D. W., et al. (1996). The Visual and Infrared Mapping Spectrometer for Cassini. In: Horn, L. (Ed.). Society of Photo-Optical Instrumentation Engineers (SPIE) Conference Series, 2803, 206–220.

Momary, T. W. and Baines, K. H. (2014). The anticyclonic eyte of the storm. Evolution of Saturn's Great Storm region and associated anticyclone as seen by Cassini/VIMS. Abstract 422.11. Proceedings of the 46[th] Annual Meeting of the Division of Planetary Sciences.

Moses, J. I., Armstrong, E. S., Fletcher, L. N., Irwin, P. G. J., Hesman, B. E., and Romani, P. N. (2015). Evolution of stratospheric chemistry in the Saturn storm beacon. Icarus 261, 149–168

Muñoz, O., Moreno, F., Molina, A., Grodent, D., Gérard, J. C., and Dols, V. (2004). Study of the vertical structure of Saturn's atmosphere using HST/WFPC2 images. *Icarus*, **169**, 413–428. Doi: 10.1016/j.icarus.2003.12.018.

Orton, G. S., Fletcher, L. N., Fouchet, T., et al. (2013). Ground-Based Observations of the Aftermath of the 2010-2011 Great Northern Springtime Storm in Saturn (Invited). *AGU Fall Meeting Abstracts*, Dec., A6.

Pedlosky J. (1979). Geophysical Fluid Dynamics, Springer-Verlag, New York, pp. 624.

Pérez-Hoyos, S., Sánchez-Lavega, A., French, R.G., and Rojas, J.F., (2005). Saturn's cloud structure and temporal evolution from ten years of Hubble space telescope images (1994–2003). *Icarus*, **176**, 155–174. Doi: 10.1016/j.icarus.2005.01.014

Pérez-Hoyos S., Sánchez-Lavega A., Solar flux in Saturn's atmosphere: maximum penetration and heating rates in the aerosol and cloud layers, ***Icarus*, 180,** 368-378 (2006).

Pierce, A.D., and Coroniti, S.C. (1966). A mechanism for the generation of acoustic-gravity waves during thunderstorm formation. *Nature*, **210**, 1209-1210. Doi: 10.1038/2101209a0.

Porco, C.C., West, R.A., Squyres, S., et al. (2004). Cassini Imaging Science: Instrument characteristics and anticipated scientific investigations at Saturn, *Space Sci. Rev.*, **115**, 363-497.

Porco, C.C., Baker, E., Barbara, J., et al. (2005). Cassini Imaging Science: Initial results on Saturn's atmosphere, *Science*, **307**, 1243-1247. Doi: 10.1126/science.1107691.

Sánchez-Lavega A. (1982). Motions in Saturn's Atmosphere: Observations before Voyager Encounters. *Icarus*, **49**, 1 - 16. Doi: 10.1016/0019-1035(82)90052-5.

Sánchez-Lavega, A., Battaner, E. (1987). The nature of Saturn's Great White Spots. *Astronomy & Astrophysics*, **185**, 315-326.

Sánchez Lavega A., Colas, F., Lecacheux, J., Laques, P., Miyazaki, I., Parker, D. (1991). The Great White Spot and disturbances in Saturn's equatorial atmosphere during 1990, *Nature*, **353**, 397 - 401. Doi: 10.1038/353397a0.

Sánchez Lavega A., Lecacheux, J., Colas, F., Laques, P. (1993). Temporal behavior of cloud morphologies and motions in Saturn's atmosphere. *Journal of Geophysical Research*, **98**, 18857 - 18872. Doi: 10.1029/93JE01777.

Sánchez-Lavega, A., Lecacheux, J. Colas, F., and Laques, P. (1994). Photometry of Saturn's 1990 equatorial disturbance. *Icarus*, **108**, 158–168. Doi: 10.1006/icar.1994.1048.

Sánchez-Lavega, A., 1994. Saturn's Great White Spots, *Chaos*, **4**, 341-353. Doi: 10.1063/1.166012.

Sánchez-Lavega A., and Gómez, J. M. (1996a). The South Equatorial Belt of Jupiter, I : Its life cycle. *Icarus*, **121**, 1 - 17. Doi: 10.1006/icar.1996.0067.

Sánchez Lavega A., Lecacheux, J., Gómez, J. M., et al. (1996b). Large-scale storms in Saturn's atmosphere during 1994, *Science*, **271**, 631 - 634. Doi: 10.1126/science.271.5249.631.

Sánchez-Lavega A., J.F. Rojas, P.V. Sada (2000). Saturn's zonal winds at cloud level, *Icarus*, 147, 405-420.

Sánchez-Lavega, A., Pérez-Hoyos, S., Rojas, J. F., Hueso, R., and French, R. G. (2003). A strong decrease in Saturn's equatorial jet at cloud level. *Nature*, **423**, 623–625. Doi: 10.1038/nature01653.

Sánchez-Lavega, A., Hueso, R., Pérez-Hoyos, S., Rojas , J. F., and French, R. G. (2004) Saturn's cloud morphology and zonal winds before the Cassini encounter. *Icarus*, **170**, 519-523. Doi: 10.1016/j.icarus.2004.05.002.

Sánchez-Lavega A., Orton, G. S., Hueso, R. et al. (2008). Depth of a strong jovian jet from a planetary-scale disturbance driven by storms, *Nature*, **451**, 437- 440. Doi: 10.1038/nature06533.

Sánchez-Lavega, A. (2011) An Introduction to Planetary Atmospheres, Taylor-Francis, CRC Press, Florida, pp. 629.

Sánchez-Lavega A., del Río-Gaztelurrutia, T., Hueso, R., et al. (2011) Deep winds beneath Saturn's upper clouds from a seasonal long-lived planetary-scale storm. *Nature*, **475**, 71-74. Doi: doi:10.1038/nature10203.

Sánchez-Lavega, A., del Río-Gaztelurrutia, T., Delcroix, M., et al. (2012). Ground-based observations of the long-term evolution and death of Saturn's 2010 Great White Spot. *Icarus*, **220**, 561-576, Doi: 10.1016/j.icarus.2012.05.033.

Sanz-Requena, J.F., Pérez-Hoyos, S., Sánchez-Lavega, A., et al. (2012). Cloud structure of Saturn's 2010 storm from ground-based imaging. *Icarus*, **219**, 142 – 149. Doi: 10.1016/j.icarus.2012.02.023.

Sayanagi, K. M., Morales-Juberias , R., and Ingersoll, A. P., (2010). Saturn's Northern Hemisphere Ribbon: Simulations and Comparison with the Meandering Gulf Stream. *Journal of the Atmospheric Sciences*, **67**, 2658-2678. Doi: 10.1175/2010JAS3315.1.

Sayanagi, K. M., Dyudina, U. A., Ewald, S. P., et al. (2013). Dynamics of Saturn's great storm of 2010-2011 from Cassini ISS and RPWS. *Icarus*, **223**, 460-478. Doi:10.1016/j.icarus.2012.12.013.

Sayanagi, K. S., Dyudina, U. A., Ewald, S. P., Muro, G. D., and Ingersoll, A. P. (2014). Cassini ISS observation of Saturn's String of Pearls. *Icarus*, **229**, 170-180. Doi: 10.1016/j.icarus.2013.10.032.

Shemansky, D. and Liu, X. (2012). Saturn Upper Atmospheric Structure from Cassini EUV and FUV Occultations. *Canadian Journal of Physics*, **90**, 817-831. Doi: 10.1139/p2012-036.

Showman, A. (2007) Numerical Simulations of Forced Shallow-Water Turbulence: Effects of Moist Convection on the Large-Scale Circulation of Jupiter and Saturn. *Journal of the Atmospheric Sciences*, **64**, 3132-3157. Doi: 10.1175/JAS4007.1.

Simon-Miller, A. A., Chanover, N. J., Orton, G. S., Sussman, M., Tsavaris, I. G., and Karkoschka, E., (2006). Jupiter's White Oval turns red. *Icarus*, **185**, 558-562. Doi: 10.1016/j.icarus.2006.08.002.

Sromovsky, L. A., Baines, K. H., and Fry, P. (2013). Saturn's Great Storm of 2010-2011: Evidence for ammonia and water ices from analysis of VIMS spectra. *Icarus*, **226**, 402–418. Doi: 10.1016/j.icarus.2013.05.043.

Sromovsky, L. A., Baines, K. H., and Fry, P. M. (2014). Vertical structure of Saturn lightning storms and storm-related dark ovals. *American Astronomical Society*, #46, # 511.09.

Sugiyama, K., Nakajima, K., Odaka, M., Ishiwatari, M., Kuramoto, K., Morikawa, Y., Nishizawa, S., Takahashi, Y. O., and Hayashi, Y.-Y. (2011), Intermittent cumulonimbus activity breaking the three layer cloud structure of Jupiter. *Geophys. Res. Lett.,* **38**, L13201; doi:10.1029/2011GL047878.
Sugiyama, K., Nakajima, K., Odaka, M., Kuramoto, K., and Hayashi Y.-Y. (2014). Numerical simulations of Jupiter's moist convection layer: Structure and dynamics in statistically steady states, *Icarus*, 229, 71-91. Doi: 10.1016/j.icarus.2013.10.016.

Tomasko, M. G., and Doose, L. R. (1984). Polarimetry and photometry of Saturn from Pioneer 11 Observations and constraints on the distribution and properties of cloud and aerosol particles, *Icarus*, **58**, 1-34. Doi: 10.1016/0019-1035(84)90096-4.

Vallis, G. K. (2006). Atmospheric and Ocean Fluid Dynamics, Cambridge University Press, 745.

Vasavada, A.R., Hörst, S. M., Kennedy, M. R., et al. (2005). Cassini imaging of Saturn: Southern hemisphere winds and vortices. *Journal of Geophysical Research.* **111**, E05004. Doi: 10.1029/2005JE002563.

Vasavada, A. R., and Showman, A. P. (2005). Jovian atmospheric dynamics: an update after Galileo and Cassini, *Reports on Progress in Physics*, **68**, 1935–1996. Doi: 10.1088/0034-4885/68/8/R06.

Warwick, J.W., Pearce, J. B., Evans, D. R., et al. (1981). Planetary radio astronomy observations from Voyager 1 near Saturn. *Science*, **212**, 239-243. Doi: 10.1126/science.212.4491.239.

West, R.A., Baines, K.H., Karkoschka, E., Sánchez-Lavega, A. (2009). Clouds and aerosols in Saturn's atmosphere. In: Dougherty, M.K., Esposito, L.W., Krimigis, S.M. (Eds.), Saturn from Cassini–Huygens. Springer, New York, pp. 113–159.

Westphal, J. A., Baum, W. A., Ingersoll, A. P., et al. (1992). Hubble Space Telescope observations of the 1990 equatorial disturbance on Saturn - images, albedos and limb darkening. *Icarus*, **100**, 485 – 498. Doi: 10.1016/0019-1035(92)90112-K.

Zakharenko, V., Mylostna, C., Konovalenko, A., et al. (2012). Ground-based and spacecraft observations of lightning activity on Saturn. *Planetary and Space Science*, **61**, 53-59. Doi: 10.1016/j.pss.2011.07.021.

Zarka, P., Cecconi, B., Denis, L., et al. (2006). Physical properties and detection of Saturn's lightning radio bursts, in: Planetary Radio Emissions VI, edited by H.O. Rucker, W.S. Kurth, and G. Mann, Austrian Academy of Sciences Press, Vienna, 111-122.

**Figure Captions**

**Figure 13.1.** Seasonal insolation cycle at the top of Saturn's atmosphere and location in latitude and solar longitude of the major planetary scale storms (Great White Spots, GWS) and other small regular convective events: (1) Lines: heat flux in $Wm^{-2}$; (2) Red dots GWS (years indicated) and the red circle is the 1994 case; (3) Small blue dots: small convective events (2002-2010); (4) Green dots: small convective events (1981-82, Voyager 1 and 2). Adapted from Pérez-Hoyos and Sanchez-Lavega (2006).

**Figure 13.2.** Mosaic of historical Great White Spots: (A) GWS 1876 (December 10). Drawing by T. W. Webb published in Sky and Telescope, January 1974); (B) GWS 1903 (July 16). Drawing by W. F. Denning published in Splendour of the Heavens (1925); (C) GWS 1933. Photographs: upper left from Lick Obs. (August 7, blue filter); upper right from Lowell Obs. (August, 9, yellow filter); lower from Lowell Obs. (September 5, yellow filter); (D) GWS 1960 (April 27). Drawing by A. Dollfus (Icarus, Vol. 2, 109. 1963); (E) GWS 1990 (October 2). CCD image from Pic-du-Midi Obs. (Sanchez-Lavega et al., 1991); (F) GWS 1990 (November 20). Image taken with the Hubble Space Telescope WFPC;

(G) GWS 1990 (November 17). Composite image over 360° longitude taken with the Hubble Space Telescope WFPC at 547 nm (Barnet et al., 1992).

**Figure 13.3.** Saturn's zonal wind meridional profile at Voyager times (red line, 1980-81) and Cassini epoch (dark line, 2005-11). The dots mark the GWS outbreak latitudes and the color bands the affected latitudes (see Table 13.1). The green bands in the southern latitudes are those where small convective events have been observed. Adapted from Sanchez-Lavega et al. (2012).

**Figure 13.4.** Top Panel: Cylindrical map (equirectangular projection) of the northern mid-latitude region. Observations at VIMS 5.1 μm channel, which uses the internal heat of Saturn as the primary lighting source, show deep features near the 3-bar level, including the thermally-bright "String of Pearls" between around 40ºW and 310ºW longitudes. Observation was obtained January 13, 2008. (From Baines et al. 2015). Lower three panels: Similar view of the String of Pearls region in ISS CB2 filter (750 nm). The 23 dark spots that are aligned at 33.2N latitude comprise the the SoP. The three panels show different longitudinal sections. The observation was made on March 29, 2008. From Sayanagi et al. (2014).

**Figure 13.5.** Cloud morphology before, during and after the storm. The panel IDs denote the image acquisition dates. Each panel is a latitude–longitude projected mosaic and shows 20.5–38.9ºN planetocentric latitude. The width of the entire figure represents 200º of longitude. Mosaics are cropped to highlight the spacing between the head (marked with red triangle) and the AV (black triangle). Yellow triangles denote the longitudes of the dark ovals. Red, green and blue color channels in these composite mosaics are assigned to Cassini ISS camera's CB2 (750 nm), MT2 (727 nm), and MT3 (889 nm) filters, respectively. From Sayanagi et al. (2013).

**Figure 13.6.** Comparison of the zonal wind measurements before, during and after the great storm of 2010-2011. All images analyzed here were captured using the CB2 filter. Panel (a) shows the maximum correlation wind profiles for May 7, 2008 (solid), September 7, 2010 (dotted), January 11, 2011 (dash-dot) and August 5, 2011 (dashed). The gray line is the zonal mean wind measurement by Sánchez-Lavega et al. (2000). Panel (b) is the same except that it enlarges the latitudes affected by the storm to highlight the wind speed change. The plus (+) symbol marks the latitude and propagation speed of the storm's head. Panel (c) shows the wind change compared to the May 7, 2008 measurement, illustrating that the measurement after the storm on August 5, 2011 shows a faster wind to the north of the storm and a slower wind to the south. From Sayanagi et al. (2013).

**Figure 13.7.** Detailed cloud properties modeling of the region of the head of the GWS in February 2011, based on Cassini ISS data. (a) Cloud top pressures ranging from 1.4 bar at the undisturbed regions to 400 mbar at the front of the disturbance. (b) Single scattering albedo at blue wavelengths (BL1) showing a dramatic increase of particle reflecivity at the compression wave formed in the front of the storm. Adapted from García-Melendo et al., (2013).

**Figure 13.8.** (A-C) VIMS spectral images of the storm in February 2011, sampled at wavelengths indicated within each panel, remapped to an `equirectangular` projection, and linearly stretched so that white corresponds to a reflectivity (I/F) of 0.4, except for D in which white corresponds to I/F=1. The parts of the storm that are bright at 4.081 μm are much darker at 3.048 μm than the region ahead of the storm. 1.887 μm and 4.081 μm are pseudo-continuum wavelengths, subject to little gas absorption, and thus show reflective properties of clouds without overlying atmospheric absorption. The strong reflectivity at 4.081 μm indicates that the cloud particles are relatively large (1 μm or more). The strong absorption in the gray region near 3.05 μm is due to particles in the storm cloud and not atmospheric gases. Adapted from Fig. 2 of Sromovsky et al. (2013).

**Figure 13.9.** (**A**) A storm head spectrum and fits to that spectrum using a main cloud layer consisting of either conservative particles (red), pure $NH_4SH$ particles (green), pure $NH_3$ particles (blue), or pure $H_2O$ particles (cyan), or pure $N_2H_4$ particles (magenta). The vertical gray bars indicate regions excluded from the analysis due to calibration uncertainties near order sorting filter joints. Relative fit quality for each entire pure VIMS spectrum (excluding the two regions indicated by vertical gray bars) is indicated by the $\chi^2/N$ (TOT) values, and for the 2.65 μm - 3.2 μm region of particulate absorption by $\chi^2/N$ (ABS) values, where there are 33 points in the ABS region and 223 in total. The yellow region between the conservative particle spectrum (red) and the VIMS measured spectrum is due to cloud particle absorption. (**B**) As in A except that the blue model spectrum is a linear combination of pure spectra, with fractions adjusted to minimize $\chi^2$ in the particle absorbing (ABS) region, and the red spectrum is a multi-layer horizontally homogeneous model spectrum with the main upper tropospheric haze layer containing an upper sub layer of conservative particles and a lower sub layer of ammonia coated water ice particles. Figure from Sromovsky et al. (2013).

**Figure 13.10.** Montage showing VLT thermal infrared images of Saturn's tropospheric storm temperatures at 18.7 and 10.7 μm, compared to a visible-light image by T. Barry, all acquired on January 19, 2011. The bottom panels B and C show the direct correspondence between the brightness temperatures (sensitive to approximately 200 mbar) and the cloud albedo variations, highlighting the warm temperatures over the western storm head, the cold-core anticyclone and the two extended cool tails to the east. The latitudinal peaks of the prograde and retrograde jets are also indicated by arrows. Adapted from Fletcher et al. (2011).

**Figure 13.11.** Evolution of the stratospheric vortex over two years as measured by Cassini/CIRS, updated from the analysis of Fletcher et al. (2012) which stopped at day 500. The upper panel shows the System III west longitude of the vortex centre, the lower panel shows the peak 7.7-μm brightness temperatures (i.e., the centre of the vortex). Different colored points represent the data source - either Cassini (black), the VLT (red) or IRTF (blue). Linear regression lines are shown as solid lines with dashed uncertainties. The green line in the top panel is the longitude of the storm head as measured by Sayanagi et al. (2013). The inset VLT image shows the stratospheric beacon at 13.1 μm on July 20, 2011.

**Figure 13.12.** Retrieved stratospheric temperatures, wind magnitudes and acetylene abundance in the stratospheric vortex at 2 mbar derived from Cassini/CIRS low-spectral resolution mapping on August 21, 2011. Latitudes are planetographic and motion around the vortex is in a clockwise direction. Adapted from Fletcher et al. (2012).

**Figure 13.13.** (a) (top-left): Celeste spectra of ethylene ($C_2H_4$) measured on 3 occasions in 2011 and 2012. The red curve (May 2011) and green (July 2011) curves were measured in phase 2 and the blue (April 2012) curve was measured in phase 3. (b) (bottom-left): CIRS maps of methane and ethylene measured in May 2011 (phase 2). The maps show that the two beacons of phase 1 have been merged into one warm bright beacon in the methane band. In the ethylene region the map shows that ethylene emission exhibits latitudinal structure (Hesman et al. 2012). (c) (top-right): The $C_2$ hydrocarbon temporal response in the pre-storm atmosphere and all phases of the storm. These abundances were measured by scaling photochemical profiles to the CIRS spectra. The scale factor on the HC profiles is shown on the y-axis (Hesman et al. 2013). (d) (bottom-right): The $C_3$ and $C_4$ hydrocarbon scale factors for pre-storm and all phases of the storm. These abundances were measured by scaling photochemical profiles of these species to the CIRS spectra (Hesman et al. 2013).

**Figure 13.14.** Dynamic spectrogram of one SED episode during the GWS on February 26, 2011. The gray bar on the right side indicates the radio wave intensity in dB above the galactic background, which is displayed as a function of time over 9 hours and frequency in logarithmic scale from 1 to 16 MHz. Besides spacecraft event time (SCET), the following Cassini orbital parameters are indicated at the abscissa: Cassini distance to Saturn's center in Saturn radii ($R_S$), sub-spacecraft western longitude (Lon), Cassini planetocentric latitude (Lat), and its local time (LT in hours). The decrease of SED rate around 20:30 is due to the disappearance of the storm's head at the eastern limb of Saturn. Later, the over-horizon effect allows observing SEDs also beyond the visible horizon.

**Figure 13.15.** SED pixels per Saturn rotation as a function of time during the GWS. The numbers of observed SED pixels (single time-frequency measurements) are plotted as black bars in a logarithmic scale. The small gap around the end of January 2011 is due to a data gap when RPWS was turned off for a short time.

**Figure 13.16.** High-spatial-esolution views of the 2010-2011 GWS by the Cassini cameras on February 26, 2011. One can clearly see the head of the storm (between 110°-120° west longitude) and the large anticyclonic vortex (between 350° to 10° longitude). The flashes detected with the blue filter are indicated as white crosses. The lower panel shows the storm ~11 h after the view in the upper panel. Both panels go over ~10° in latitude and ~140° in western longitude. Adapted from Dyudina et al. (2013).

**Figure 13.17.** The head of the Great Storm GWS 2010-11 from the analysis of Cassini ISS images on February 26, 2011. (a) Wind velocity field vector superimposed to an image of the head of the storm; (b) Mean wind velocity map; (c) Vorticity map; (d) (e) Images of the head separated by 10.5 hr showing rapid changes of the cloud tops. Adapted from García-Melendo et al. (2013).

**Figure 13.18.** Simulation results of the GWS for the EPIC and SW models. Left column panels: evolution of the first 12 days (top to bottom) of the PV field according to EPIC simulations. This result was obtained for the lowest values of $N^2$ for the deepest layers and zonal winds constant or near constant and reproduces the observations quite well (compare with Fig. 13.5; see also Sanchez-Lavega et al., 2011). Right: Potential Vorticity and velocity maps according to EPIC (upper pair) and SW (lower pair) model simulations.

**Figure 13.19.** Global evolution of the storm during the first 90 days according to SW models. The three upper bottom panels represent the concentration of a passive digital "dye" introduced in the simulation to reproduce the growth and expansion of the storm. Tracer distribution is indistinguishable from the resulting PV fields. Some of the main morphological aspects of the storm are well reproduced, such as the turbulent expansion of the south branch, including chains of small vortices, the formation of the northern branch, and of a long-lived vortex at the end of the tail. The bottom panel represents the velocity field showing that the main structure of the tail is formed by anticyclonic cells at latitude 40-43°N generated by the storm Simulations were performed with a horizontal resolution of ~ 200 km and $L_R \cong 350$ km. Based on García-Melendo and Sanchez-Lavega (2016).

**Figure 13.20.** Zonal wind alterations around the latitude of the 2010 GWS onset. The grey line represents the observed zonal winds before the storm outbreak The solid line is the new zonal wind profile measured in February 2012, six months after the storm demise, showing an altered profile. The dashed line is the result of Shallow Water simulations after six months of storm evolution (starting from the grey profile), showing the winds affected by the anticyclonic vorticity injected by the storm.

**Figure 13.21.** This cartoon summarizes the authors' views of the three-dimensional structure of the Great Saturn Storm of 2010-11. From P. L. Read, Nature, 475, 44-45 (2011).

## Table 13.1: Great White Spots basic properties

| Event | First detection | Orbital longitude | Latitude Planetographic (Planetocentric) | Velocity $(ms^{-1})$ | Affected latitude band | Ambient vorticity $(s^{-1})$ | Planetary vorticity $(s^{-1})$ |
|---|---|---|---|---|---|---|---|
| GWS 1876 | 1876.9 | 170° | 8°±3°N  (6.5°) | 396 | ~ 0° – 20°N | $-4 \times 10^{-6}$ | $4.6 \times 10^{-5}$ |
| GWS 1903 | 1903.5 | 130° | 36°±2°N (30.6°) | 19 | ~ 30° – 45°N | $-7 \times 10^{-6}$ | $1.8 \times 10^{-4}$ |
| GWS 1933 | 1933.7 | 134° | 2°±3°N (1.6° | 400 | ~ -5° – 20°N | $2.0 \times 10^{-5}$ | $1.1 \times 10^{-5}$ |
| GWS 1960 | 1960.25 | 106° | 58°N ± 1° (52.5°) | 4 | 48°N – 60°N [78°N]* | $10^{-5}$ | $2.8 \times 10^{-4}$ |
| GWS 1990 | 1990.9 | 121° | 12°N ± 1° (9.8°) [5°N ±2°]† (4.1°) | 365.0 [402.0]† | 15°S – 25°N | $-4 \times 10^{-5}$ $[2.0 \times 10^{-6}]$† | $6 \times 10^{-5}$ |
| GWS 2010 | 2010.93 | 16° | 38°N (32.4°) [41°N ± 1°]& (35.2°) | -27.8 | 25°N – 48°N | $3 \times 10^{-6}$ | $1.9 \times 10^{-4}$ |

Notes: Data refer to the center of the main spot (usually known as the GWS storm head) that is assumed to be the convective source.

The latitude of a point on an equipotential surface of an oblate planet is defined by the angle between a line through the point and the equatorial plane. It is planetocentric if the line intersects the planet's center. It is planetographic if the line is normal to the equipotential surface.

For the ambient vorticity (meridional shear of the zonal wind) the positive sign means anticyclonic and the negative cyclonic. See Figure 2. The planetary vorticity is given by the Coriolis parameter *f*.

†The onset of the 1990 storm took place at 12°N, but after two weeks the stormhead center migrated towards the Equator.

* The 1960 GWS clouds expanded meridionally up to ~80N after a month of its outbreak.

& The onset of the 2010 GWS took place at a westward jet but the head centre migrated toward the North.

FIGURE 13.1

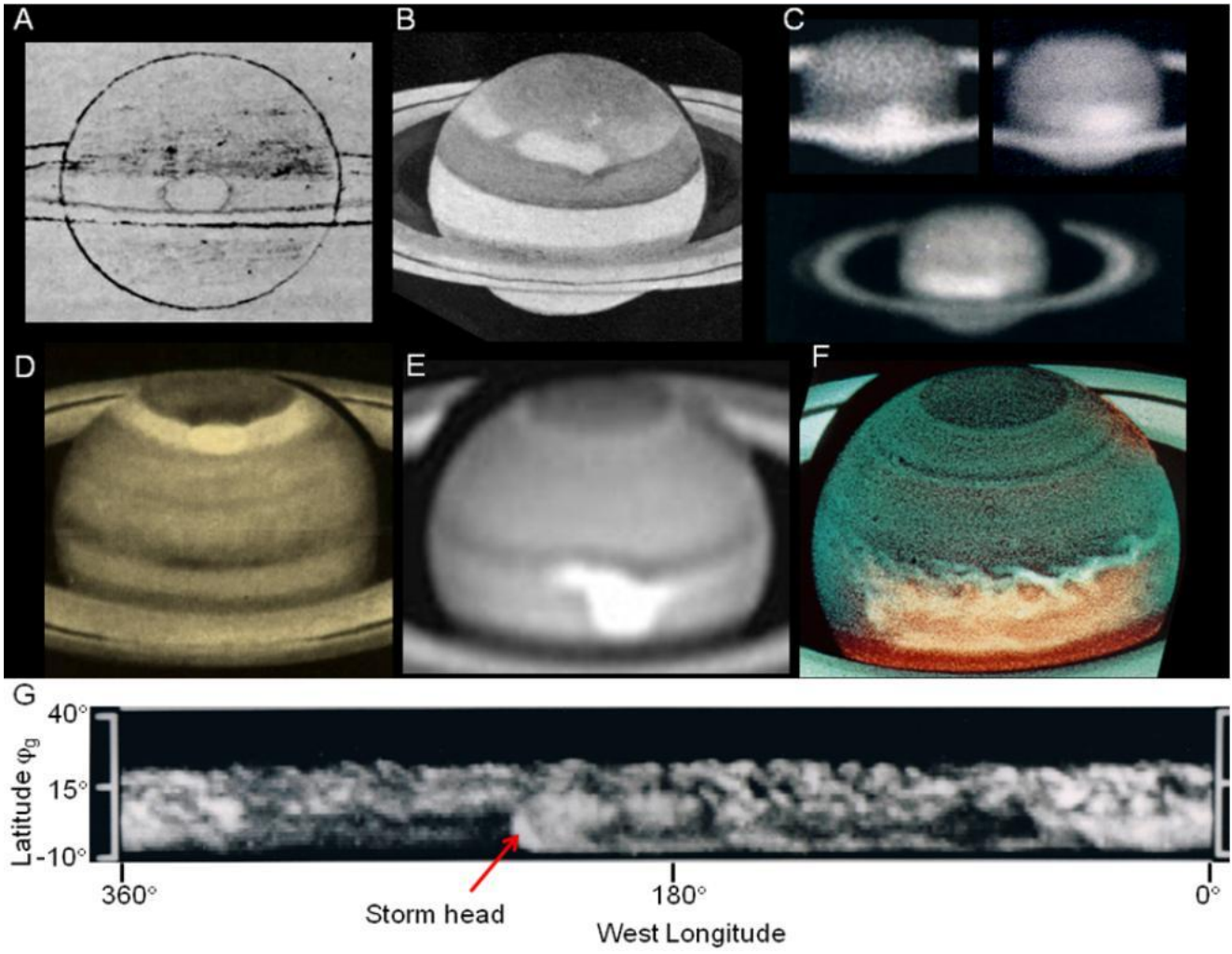

FIGURE 13.2

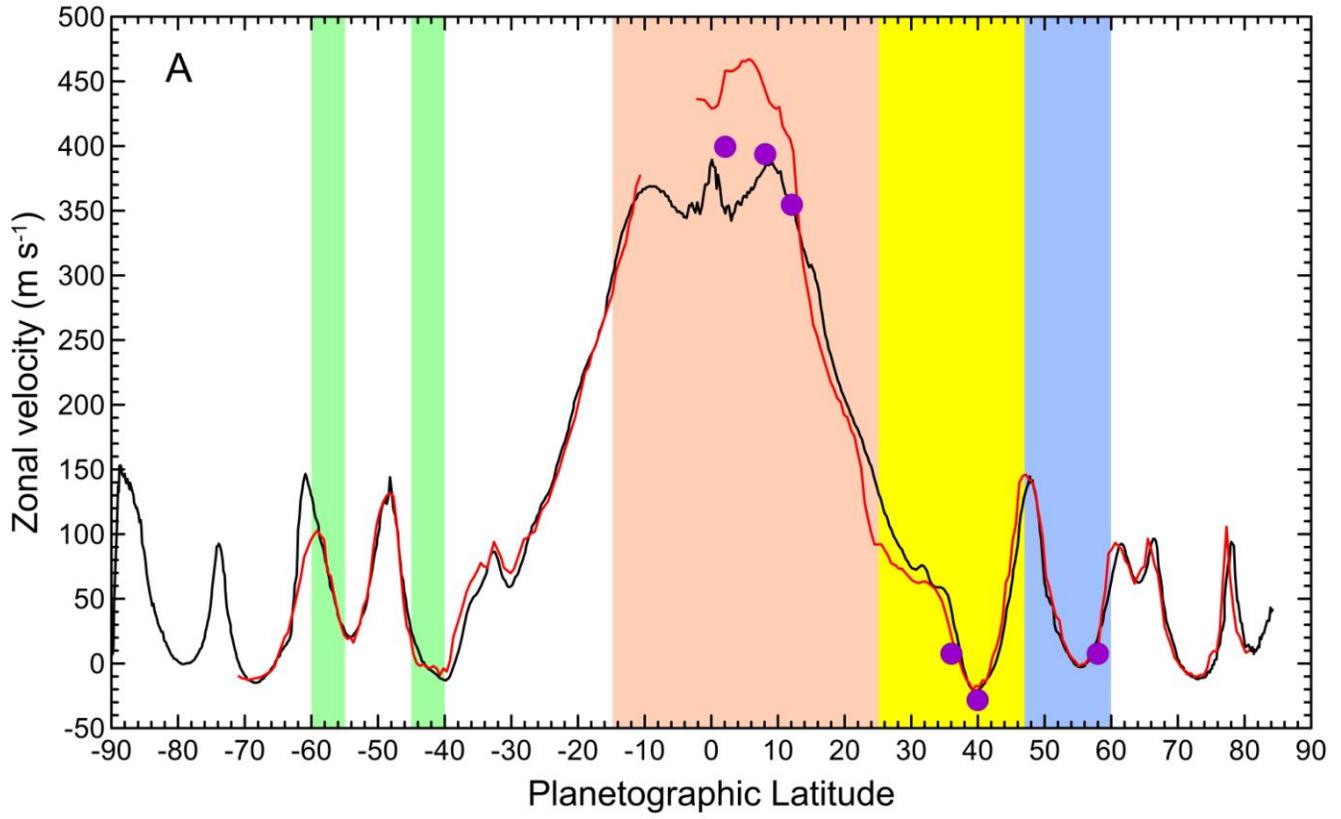

FIGURE 13.3

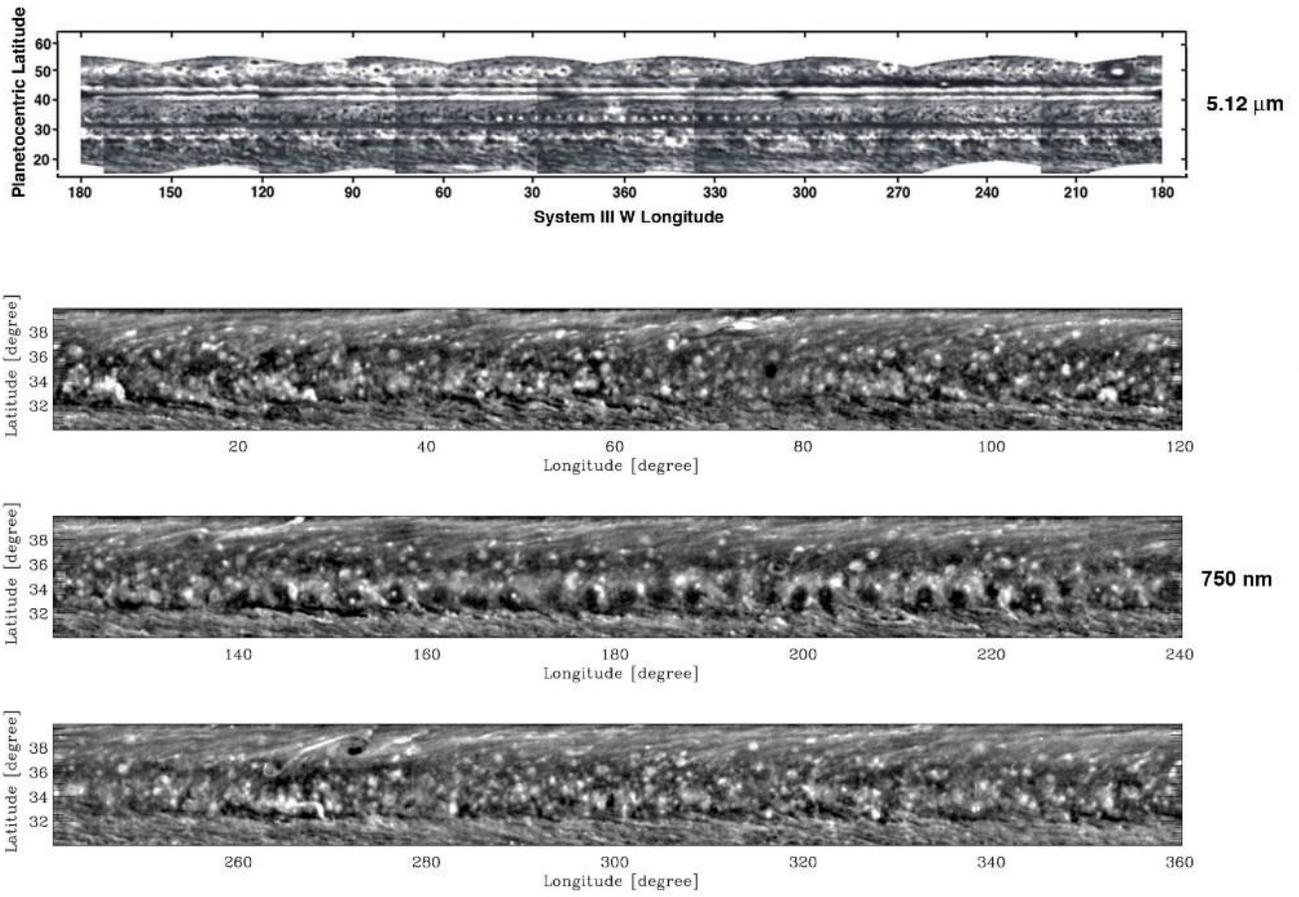

FIGURE 13.4

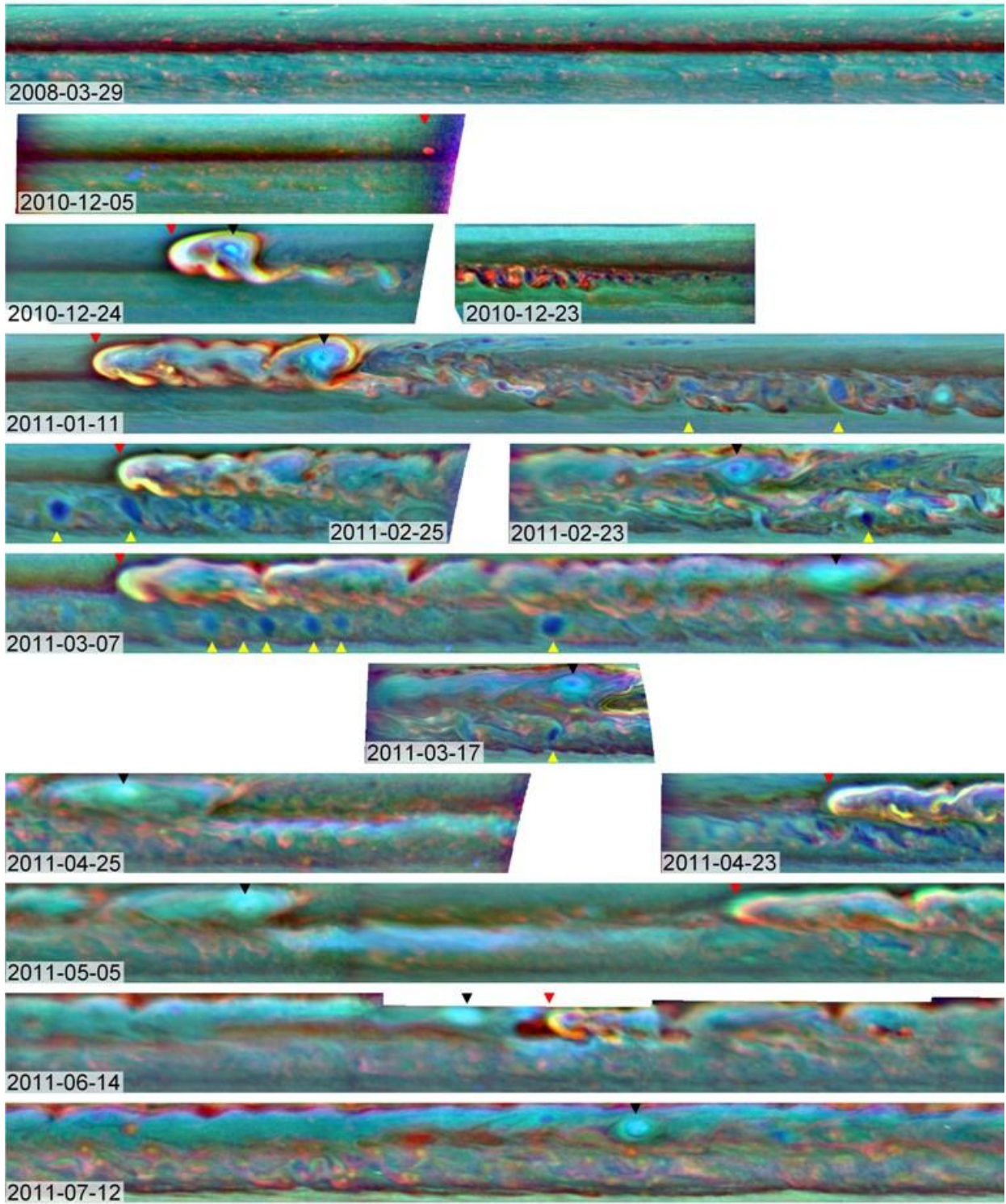

FIGURE 13.5

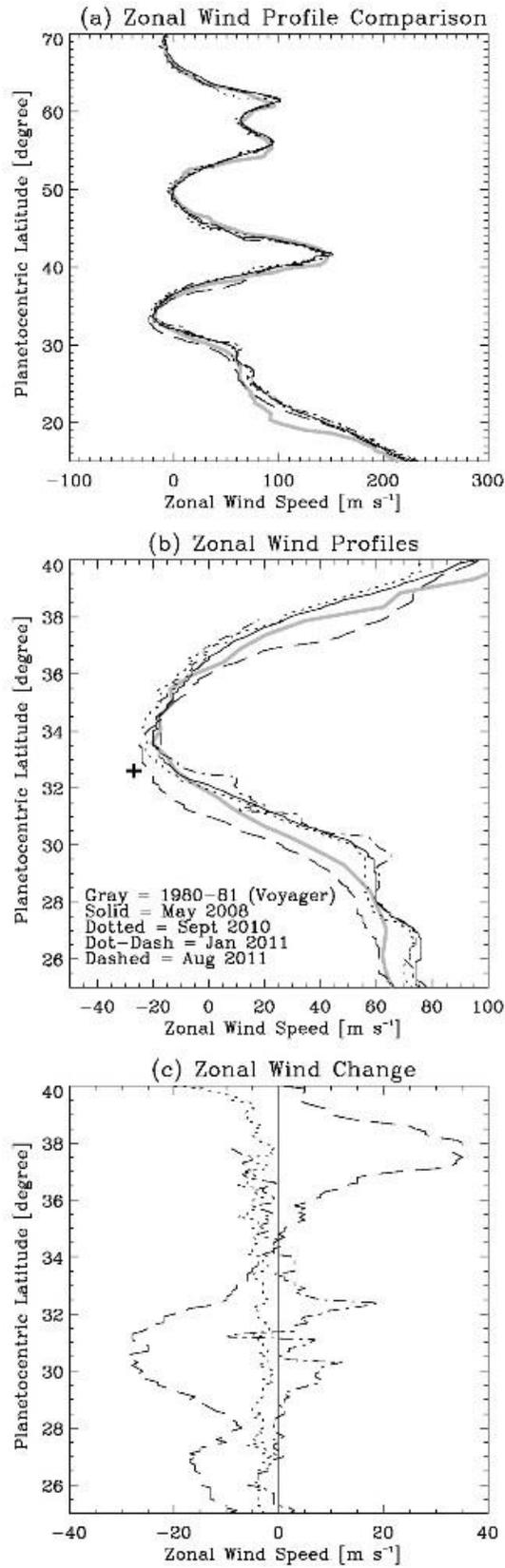

FIGURE 13.6

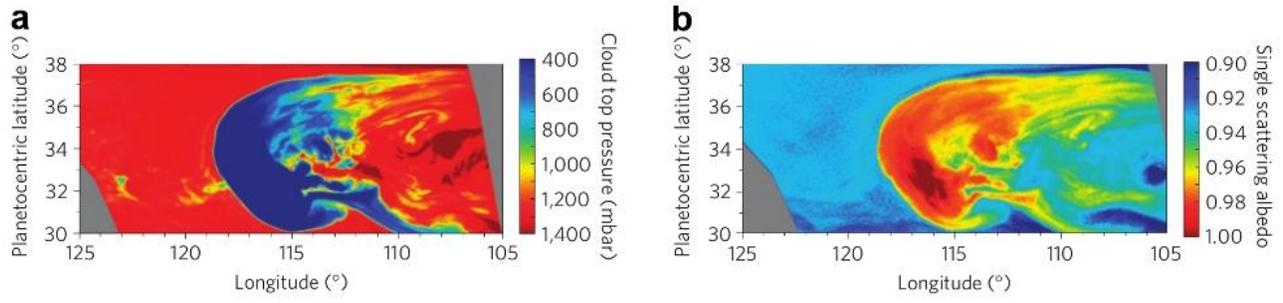

FIGURE 13.7

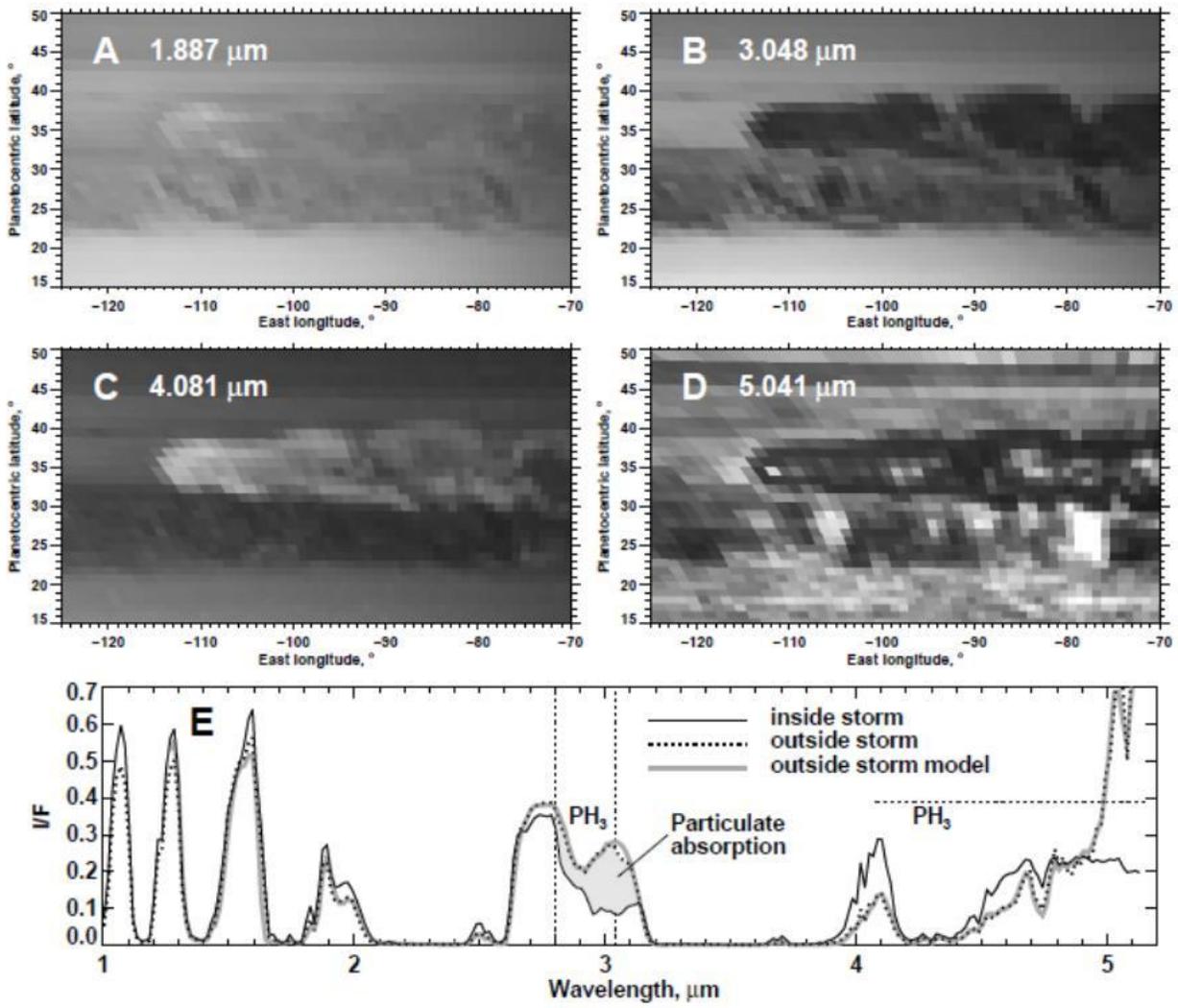

FIGURE 13.8

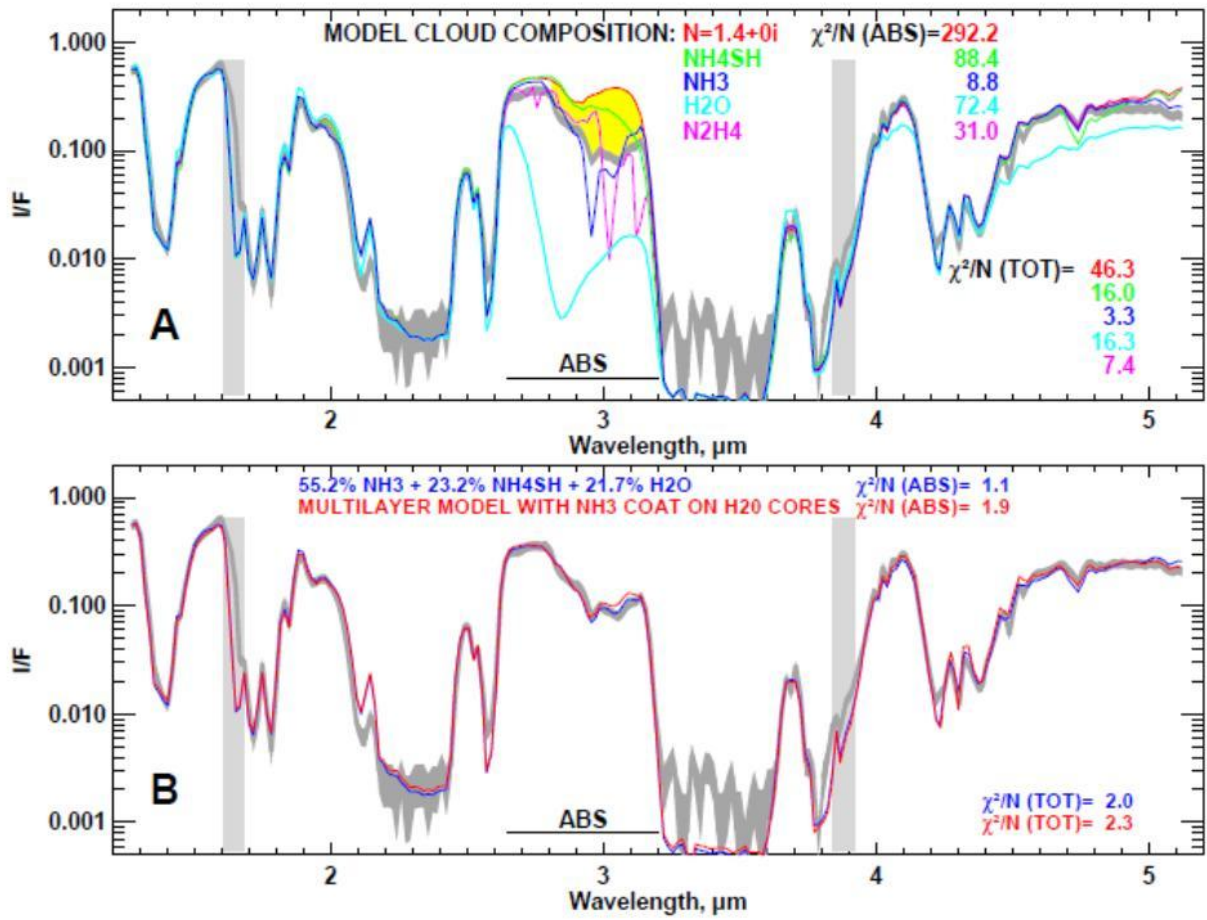

FIGURE 13.9

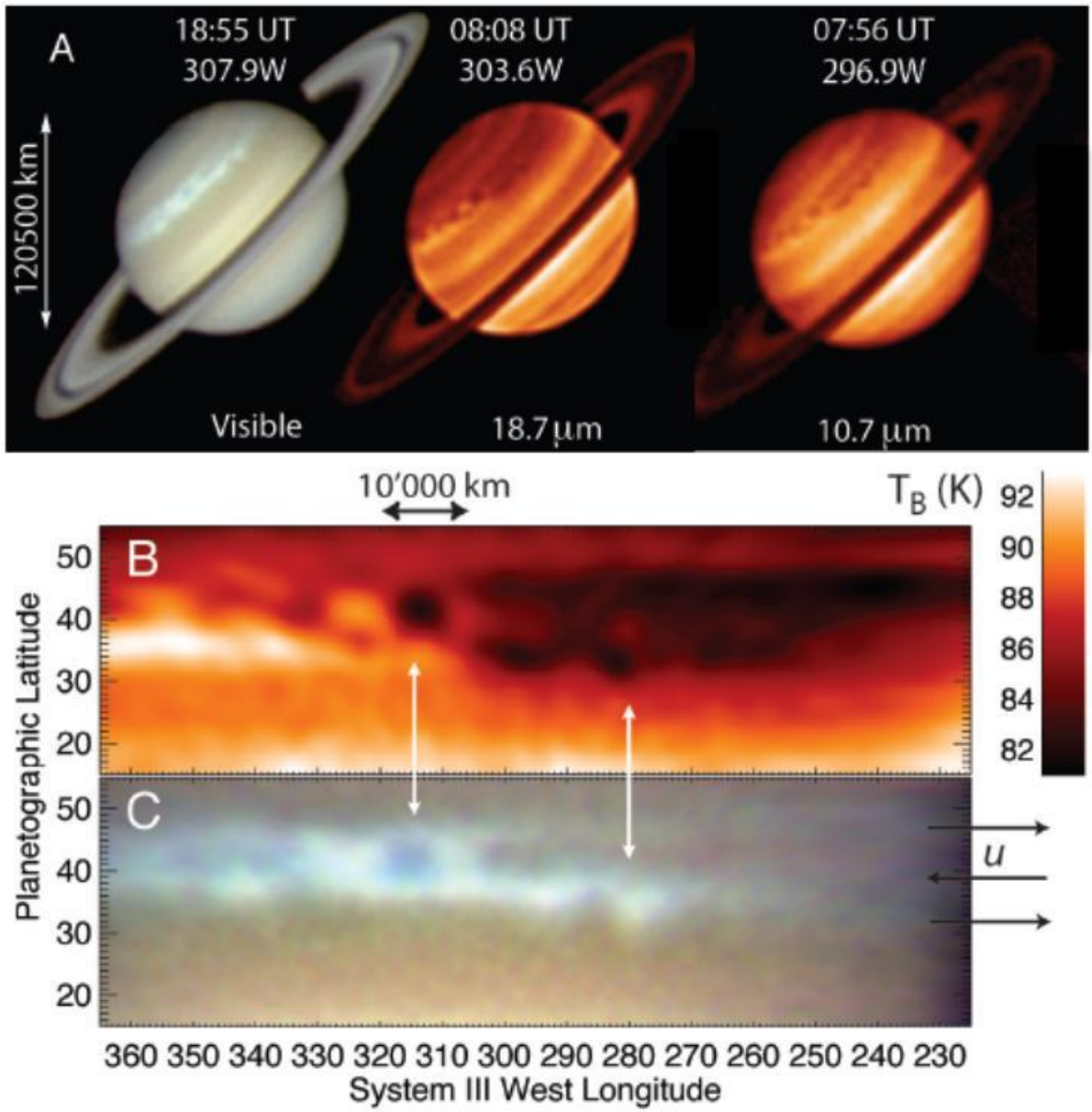

FIGURE 13.10

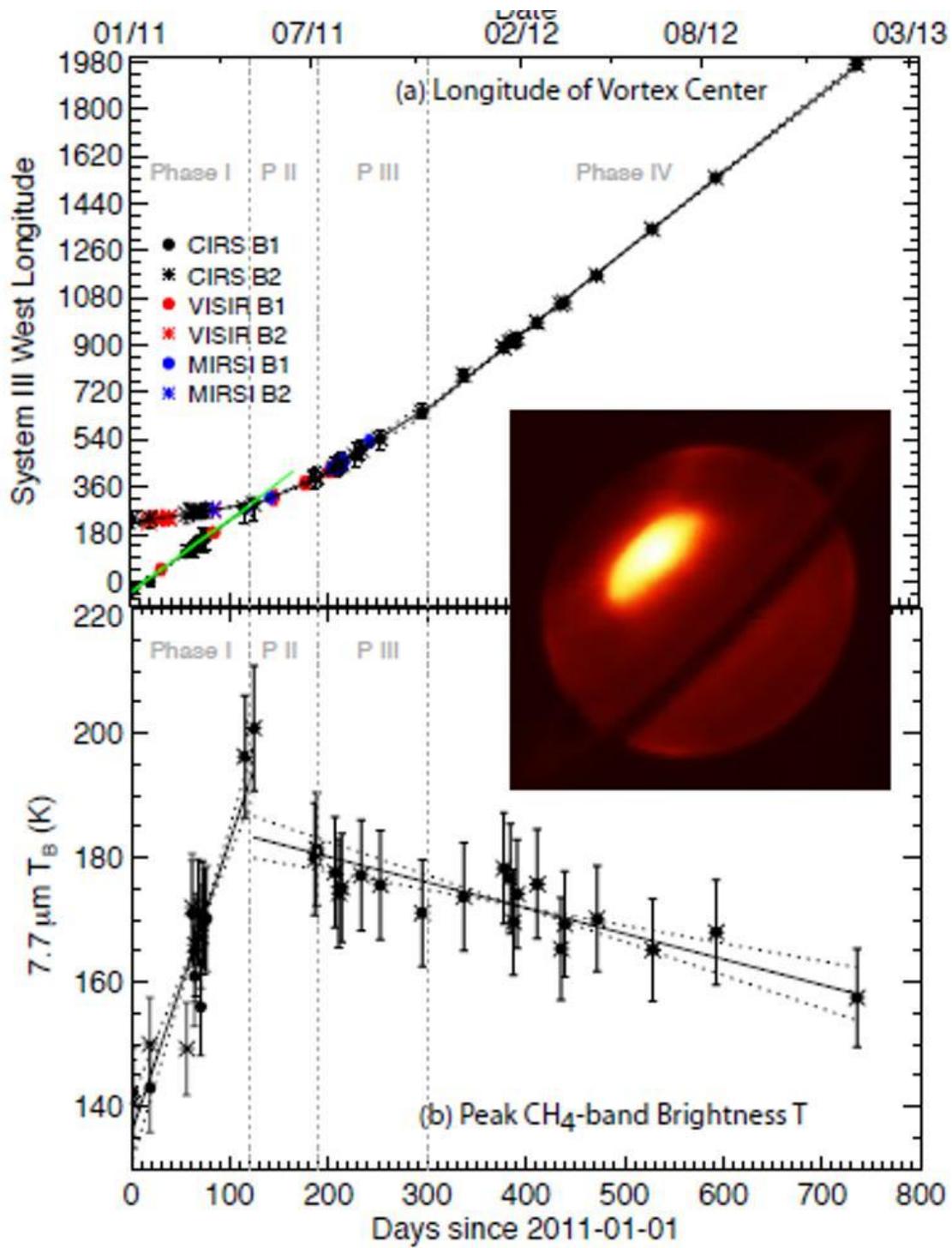

FIGURE 13.11

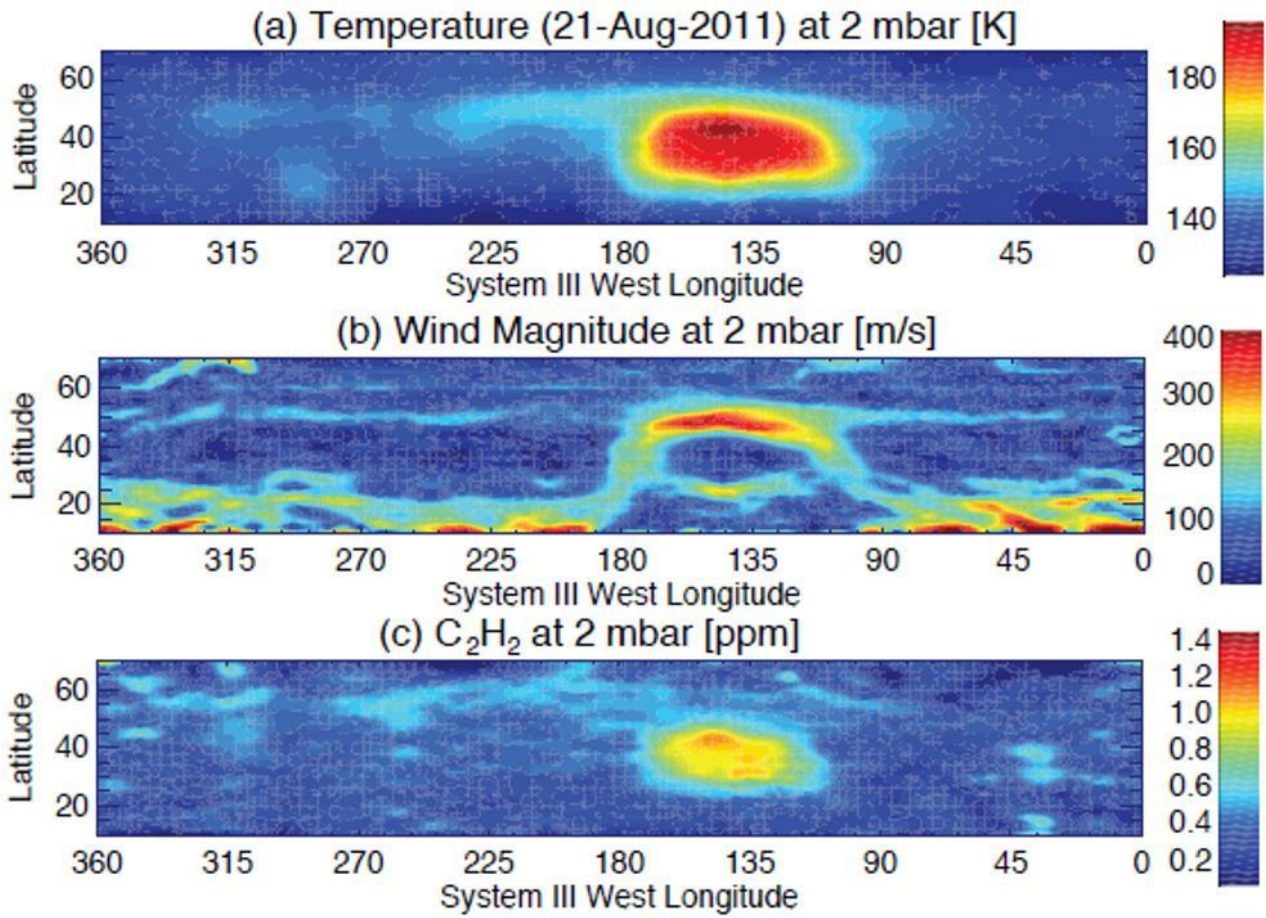

FIGURE 13.12

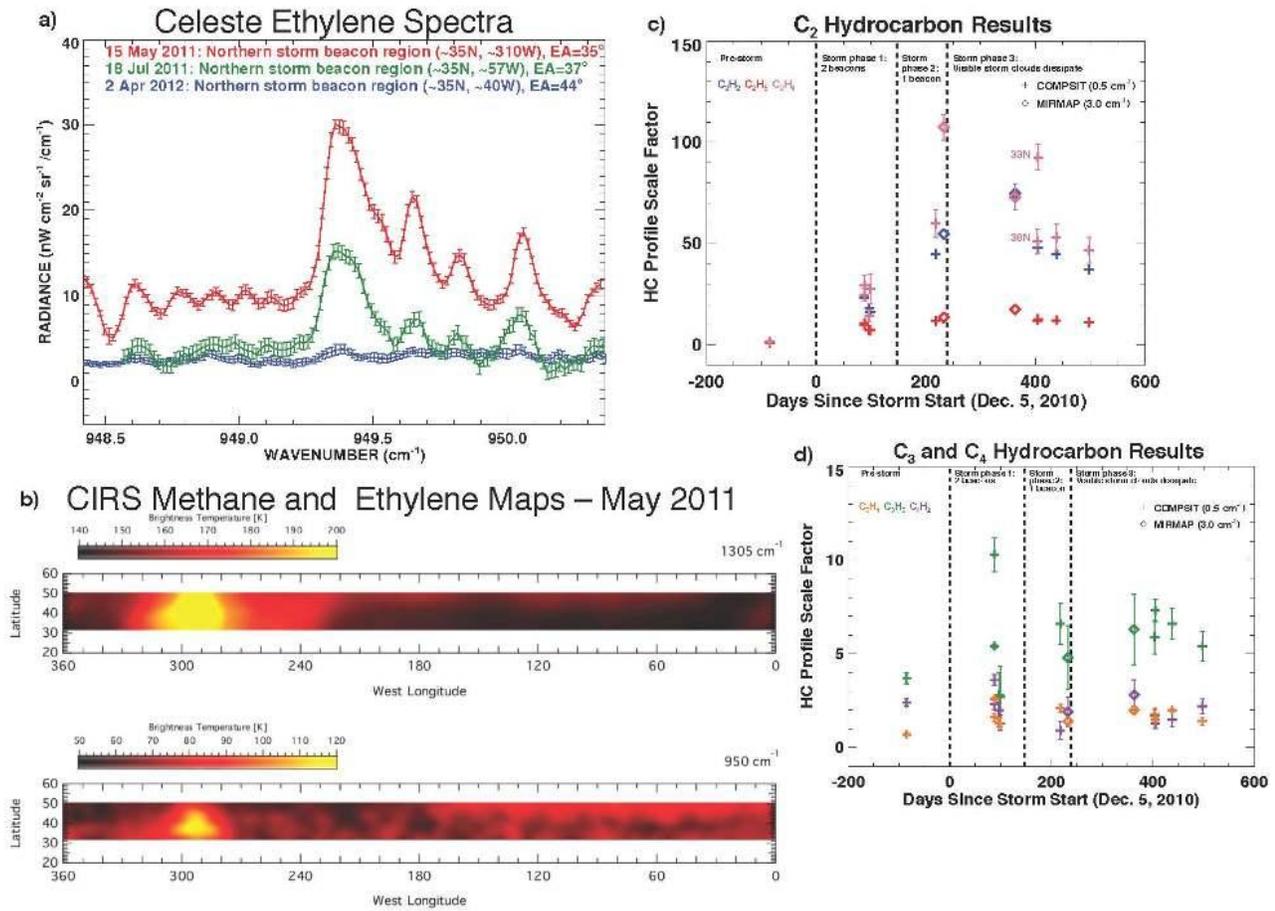

FIGURE 13.13

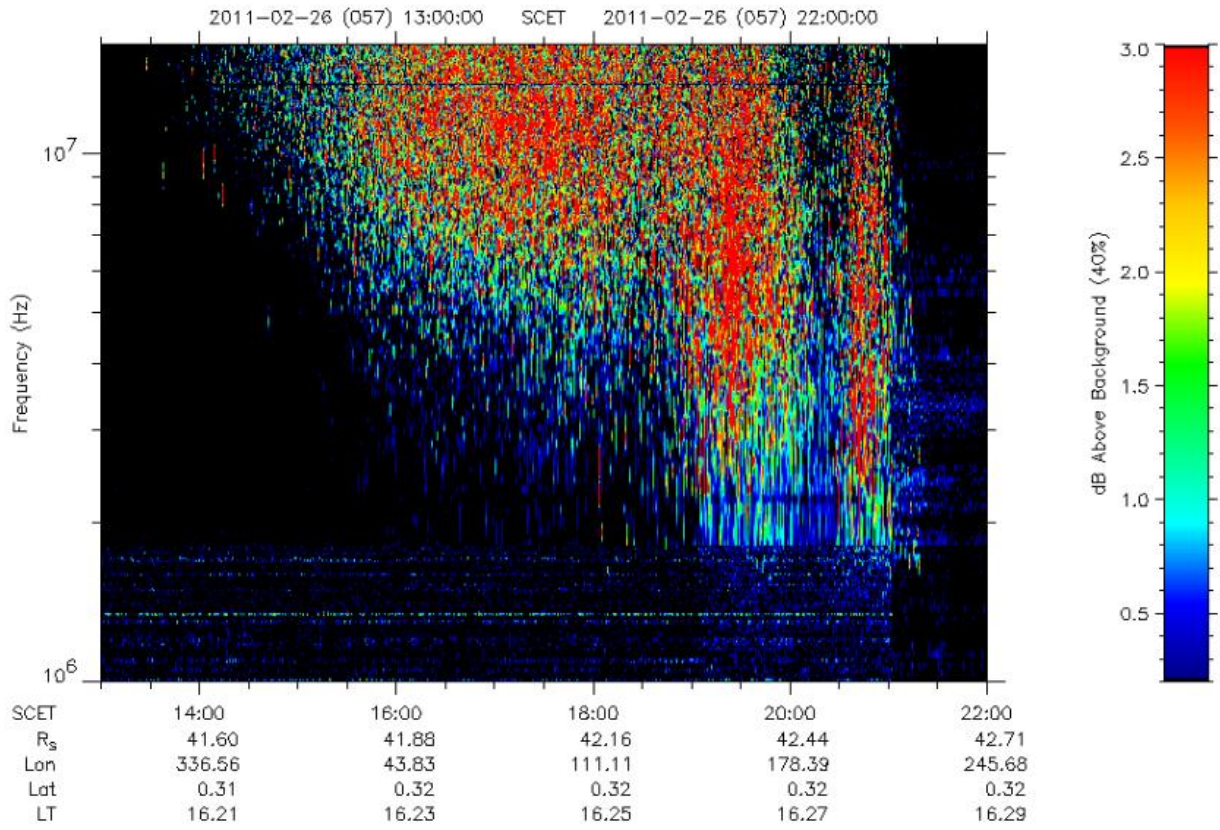

FIGURE 13.14

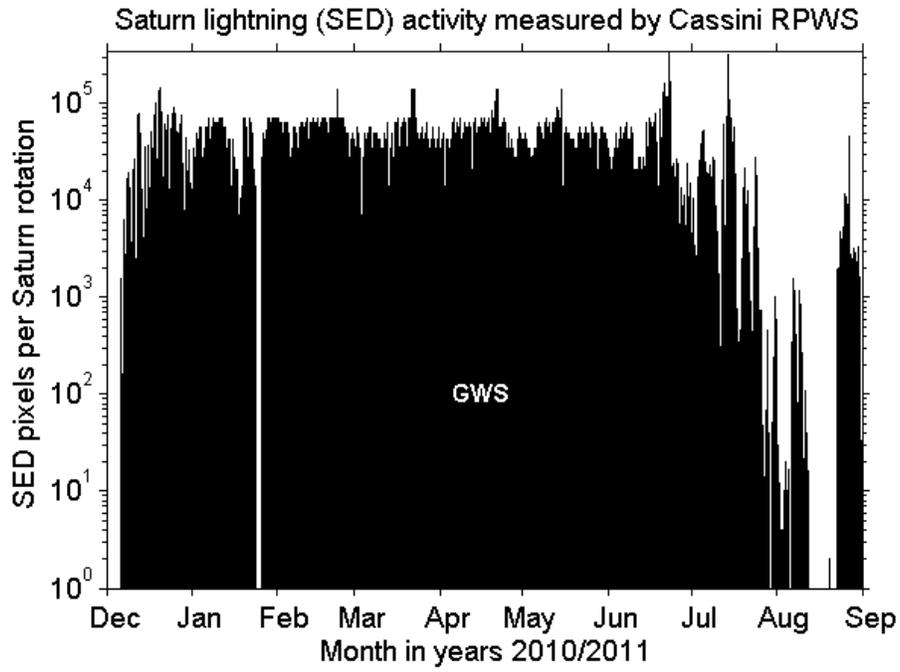

FIGURE 13.15

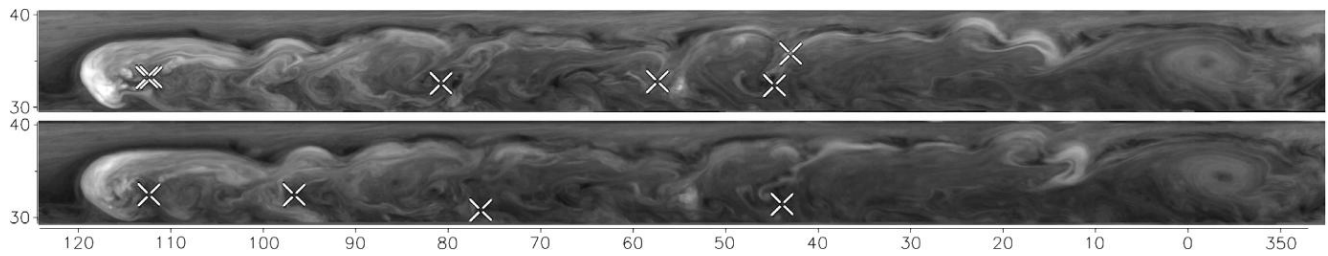

FIGURE 13.16

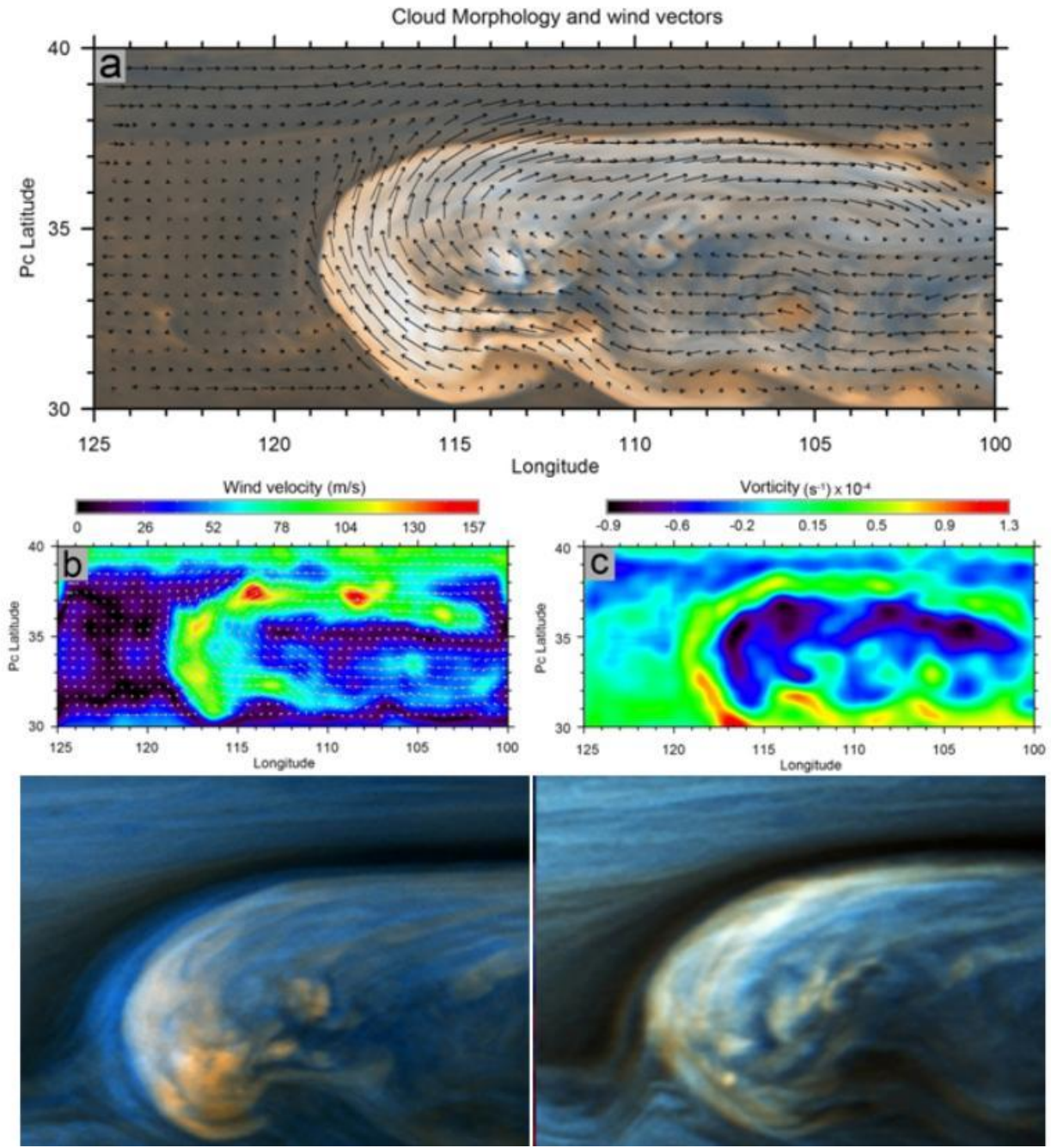

FIGURE 13.17

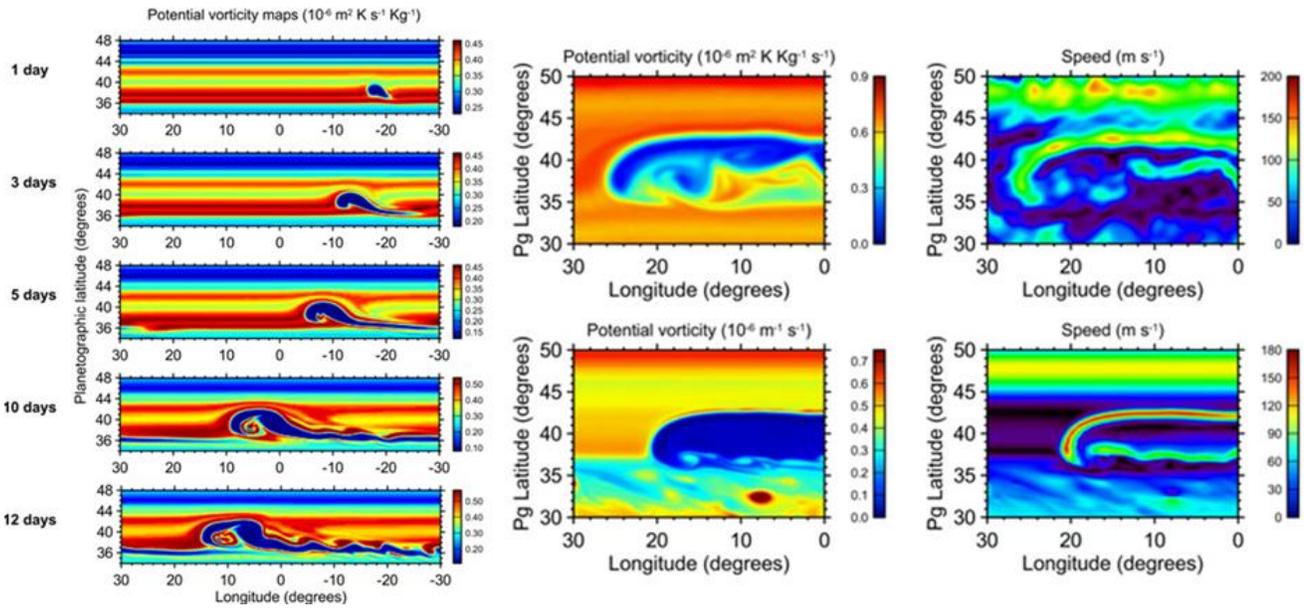

FIGURE 13.18

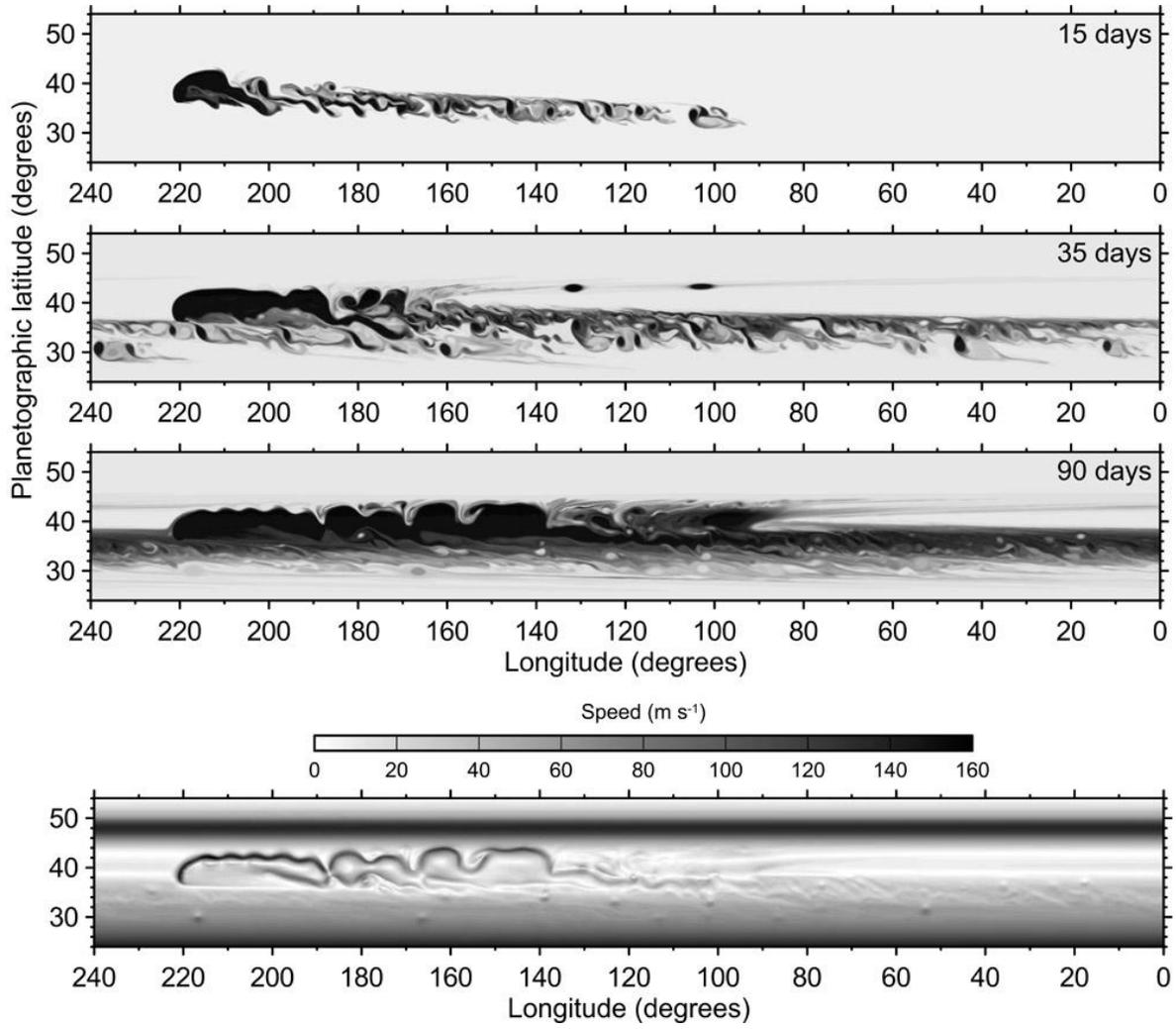

FIGURE 13.19

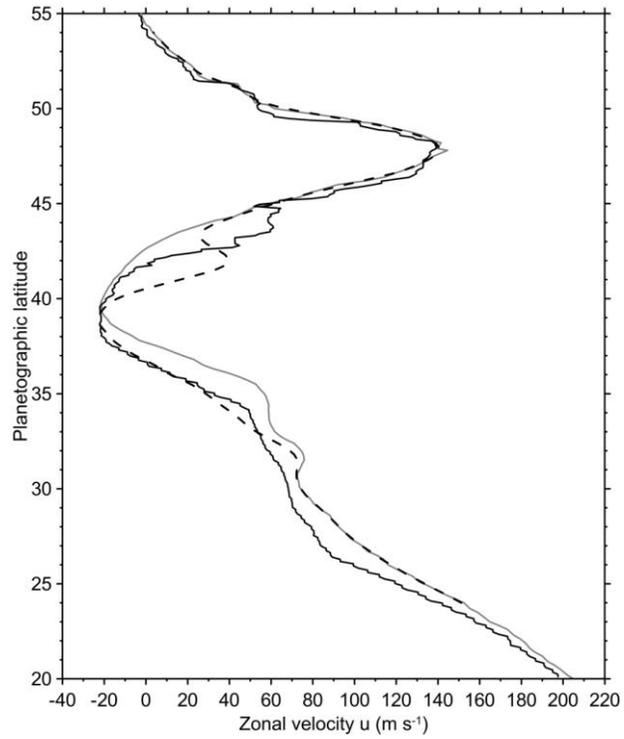

FIGURE 13.20

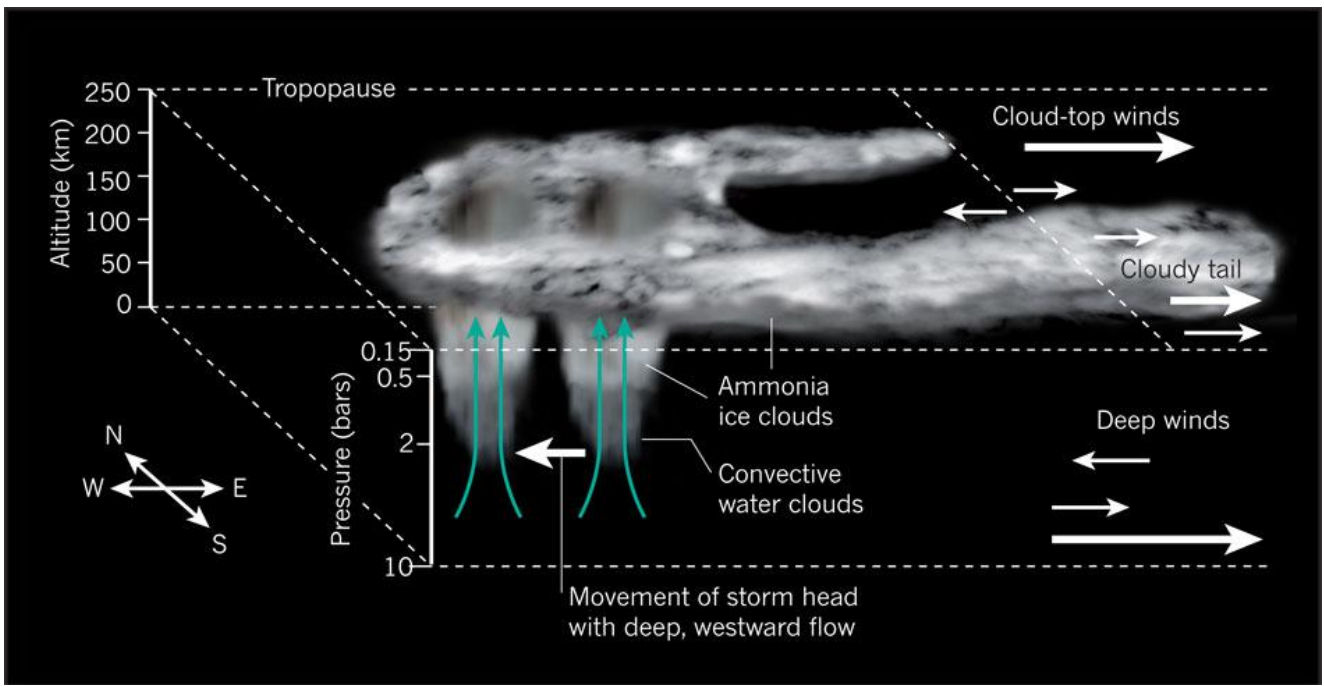

FIGURE 13.21